\documentclass[12pt,a4paper]{article}
\usepackage{graphicx}
\usepackage{subeqnarray}
\begin{document}
%
%
%
%
\def\astrobj#1{#1}
\newenvironment{lefteqnarray}{\arraycolsep=0pt\begin{eqnarray}}
{\end{eqnarray}\protect\aftergroup\ignorespaces}
\newenvironment{lefteqnarray*}{\arraycolsep=0pt\begin{eqnarray*}}
{\end{eqnarray*}\protect\aftergroup\ignorespaces}
\newenvironment{leftsubeqnarray}{\arraycolsep=0pt\begin{subeqnarray}}
{\end{subeqnarray}\protect\aftergroup\ignorespaces}
\newcommand{\diff}{{\rm\,d}}
\newcommand{\pprime}{{\prime\prime}}
\newcommand{\szeta}{\mskip 3mu /\mskip-10mu \zeta}
\newcommand{\FC}{\mskip 0mu {\rm F}\mskip-10mu{\rm C}}
\newcommand{\appleq}{\stackrel{<}{\sim}}
\newcommand{\appgeq}{\stackrel{>}{\sim}}
\newcommand{\legr}{\stackrel{<}{>}}
\newcommand{\grle}{\stackrel{>}{<}}
\newcommand{\Int}{\mathop{\rm Int}\nolimits}
\newcommand{\Nint}{\mathop{\rm Nint}\nolimits}
\newcommand{\range}{{\rm -}}
\newcommand{\displayfrac}[2]{\frac{\displaystyle #1}{\displaystyle #2}}
\def\astrobj#1{#1}
%
\title{
Globular cluster star classification: \\ application to M13}
\author{{R.~Caimmi}\footnote{
{\it Physics and Astronomy Department, Padua Univ., Vicolo Osservatorio 3/2,
I-35122 Padova, Italy}
email: roberto.caimmi@unipd.it~~~
fax: 39-049-8278212}
\phantom{agga}}
%
%
\maketitle
\begin{quotation}
\section*{}
\begin{Large}
\begin{center}
Abstract

\end{center}
\end{Large}
\begin{small}

Starting from recent determination of Fe, O, Na abundance on a restricted
sample $(N=67)$ of halo and thick disk stars, a natural and well motivated
selection criterion is defined for the classification of globular cluster
stars.   An application is performed to M13 using a sample $(N=112)$ for
which Fe, O, Na abundance has been recently inferred from observations.
A comparison is made between current and earlier M13 star classification.
Both O and Na empirical differential abundance distributions are
determined for each class and the whole sample (with the addition of Fe in the
last case) and compared with their theoretical counterparts due to cosmic
scatter obeying a Gaussian distribution whose parameters are inferred from
related subsamples.   The occurrence of a fit between empirical and
theoretical distribution is interpreted as absence of significant chemical
evolution and vice versa.   The procedure is repeated with regard to four
additional classes according if oxygen and sodium abundance is above (stage
CE) or below (stage AF) a selected threshold.   Both O and Na empirical
differential abundance distributions, related to the whole sample, exhibit a
linear fit for AF and CE stage.   Within the errors, the oxygen slope for CE
stage is equal and opposite in sign with respect of the sodium slope for AF
stage, while the contrary holds in dealing with the oxygen slope for AF stage
with respect to the sodium slope for CE stage.   In the light of simple models
of chemical evolution applied to M13, oxygen depletion appears to be mainly
turned into sodium enrichment for [O/H]$\ge-1.35$ and [Na/H]$\le-1.45$, while
one or more largely preferred channels occur for [O/H]$<-1.35$ and
[Na/H]$>-1.45$.   In addition, the primordial to current M13 mass ratio can be
inferred from the true sodium yield in units of the sodium solar abundance.
Though the above results are mainly qualitative due to large $(\mp1.5$ dex)
uncertainties in abundance determination, still the trend exhibited is
expected to be real.    The proposed classification of globular cluster stars
may be extended in a twofold manner, namely to (i) elements other than Na and
Fe and (ii) globular clusters other than M13.

\noindent
{\it keywords - 
Galaxy: evolution - Galaxy: formation - Galaxy: halo - globular clusters:
general - globular clusters: individual (M13).}
\end{small}
\end{quotation}

\section{Introduction} \label{intro}

\noindent\noindent
Globular clusters (GCs) are fundamental building blocks of galaxies and most
of them are among the oldest stellar systems.   Abundance analysis of GC stars
provides valuable clues for understanding the evolution of both the cluster in
itself and the hosting galaxy.

In the past, GCs were conceived as the result of an initial burst of highly
efficient star formation, where the remaining gas was blown up by type II
supernova (SNII) explosions together with SNII ejecta.   Gas returned later
from planetary nebulae was blown up by type Ia supernova (SNIa) explosions
together with SNIa ejecta.   Accordingly, GC stars were expected to be coeval
and with element abundance affected only by cosmic scatter.

Abundance surveys with increasingly precise instrumentation disclosed
consistent star-to-star abundance variation of light elements (from C to Al).
At least in several cases, significant
variations in He abundance were inferred.   For further details and complete
references, an interested reader is addressed to recent comprehensive GC
abundance surveys (e.g., Carretta et al. 2009a, 2009b), reviews (e.g., Piotto
2009; Gratton et al. 2012) and investigations (e.g., Johnson and Pilachowski
2012, hereafter quoted as JP12; Conroy 2012, hereafter quoted as C12).

More specifically, interpretation of recent data disclosed the following. (i)
GC normal (i.e. similar photospheric composition with respect to field halo
stars) and anomalous (i.e. different photospheric composition with respect to
field halo stars) stars affect the total mass to a comparable extent. (ii)
Light element abundance undergoes a continuous variation passing from GC
normal to most anomalous stars. (iii) GC anomalous stars exhibit enhanced N,
Na, Al; depleted O, Mg; more or less unchanged Si, Ca, Fe; enhanced He;
anticorrelations such as O-Na, Mg-Al; multiple evolutionary sequences in the
colour-magnitude diagram.

An explanation of the above mentioned items, within the framework of a single
model, is not an easy matter.   For instance, GC normal and anomalous stars in
comparable proportion would imply more massive GCs at birth and/or
substantially different initial stellar mass function if anomalous abundances
are due to asymptotic giant branch (AGB) stars $(3\appleq m/m_\odot\appleq8)$;
a continuous variation of light elements passing from GC normal to most
anomalous stars would imply (at least) two star generations separated by a
time interval larger than about 0.1 Gyr and, in addition, inhomogeneous
mixing between recycled material from AGB stars and inflowing primordial gas,
or be a mere effect of mesaurement errors;
O-Na and Mg-Al anticorrelations together with enhanced He would imply
high-temperature proton-capture burning within AGB stars and/or rapidly
rotating massive main-sequence stars and/or massive binary stars; for further
details and complete references an interested reader is addressed to recent
investigations (e.g., JP12; C12) and proceedings (e.g., Renzini 2013; Ventura
et al. 2013).

GC anomalous stars are usually subclassified as extreme and intermediate,
according if light elements are substantially or moderately enhanced/depleted
with respect to field halo stars of similar Fe abundance (e.g., JP12).   On
the other hand, no general consensus still exists about a definition of GC
normal (or primordial), intermediate and extreme stars, in absence of a
related physical criterion for distinguishing one from the others.   A
rigorous GC star classification would be useful for tracing the past history
of the Galaxy and, in fact, can be made using recent abundance determinations
available for a sample of halo and low-metallicity ([Fe/H] $<-0.6$) thick disk
stars (Nissen and Schuster 2010,
hereafter quoted as NS10; Ramirez et al. 2012, hereafter quoted as Ra12).

The present note is aimed to this respect, where special effort is devoted to
O, Na, Fe, whose abundances 
are known for about one hundred M13 stars (JP12).
Following a current attempt (Caimmi 2013), a rigorous GC star classification
is provided in Section 2.   An application to M13, including both star
classification and differential element abundance distribution, is shown in
Section 3.   The implications for halo formation and evolution, in the light
of a simple model of chemical evolution, are discussed in Section 4.   The
conclusion is outlined in Section 5.

\section{A rigorous GC star classification} \label{GCsc}

The fractional logarithmic number abundance or, in short, number abundance,
can be inferred from recently studied samples of solar neighbourhood FGK-type
dwarf stars (NS10; Ra12), as [Q/H]$=$[Q/Fe]$-$[Fe/H], for Q = O, Na, Mg, Si,
Ca, Ti, Cr, Ni, Fe.   More specifically, sample (HK) stars can be subsampled
as low-$\alpha$ halo (LH), high-$\alpha$ halo (HH), thick disk (KD), GC
outliers (OL).   The related population is $N=67$ (HK), 24 (LH), 25 (HH), 16
(KD), 2 (OL).   Oxygen can be taken as reference element in that it is the
most abundant metal and, in addition, mainly synthesised within SNII
progenitors.

With regard to the $({\sf O}$\,[O/H]\,[Q/H]) plane, stars belonging to
different environments display along a ``main sequence'', [Q,O] =
$[a_{\rm Q},b_{\rm Q},\Delta b_{\rm Q}]$, bounded by two parallel straight
lines:
\begin{lefteqnarray}
\label{eq:QOH}
&& [{\rm Q/H}]=a_{\rm Q}[{\rm O/H}]+b_{\rm Q}\mp\frac12\Delta b_{\rm Q}~~;
\end{lefteqnarray}
with the possible exception of OL stars.
For further details refer to a current attempt (Caimmi 2013).

\begin{figure*}[t]
\begin{center}      
\includegraphics[scale=0.8]{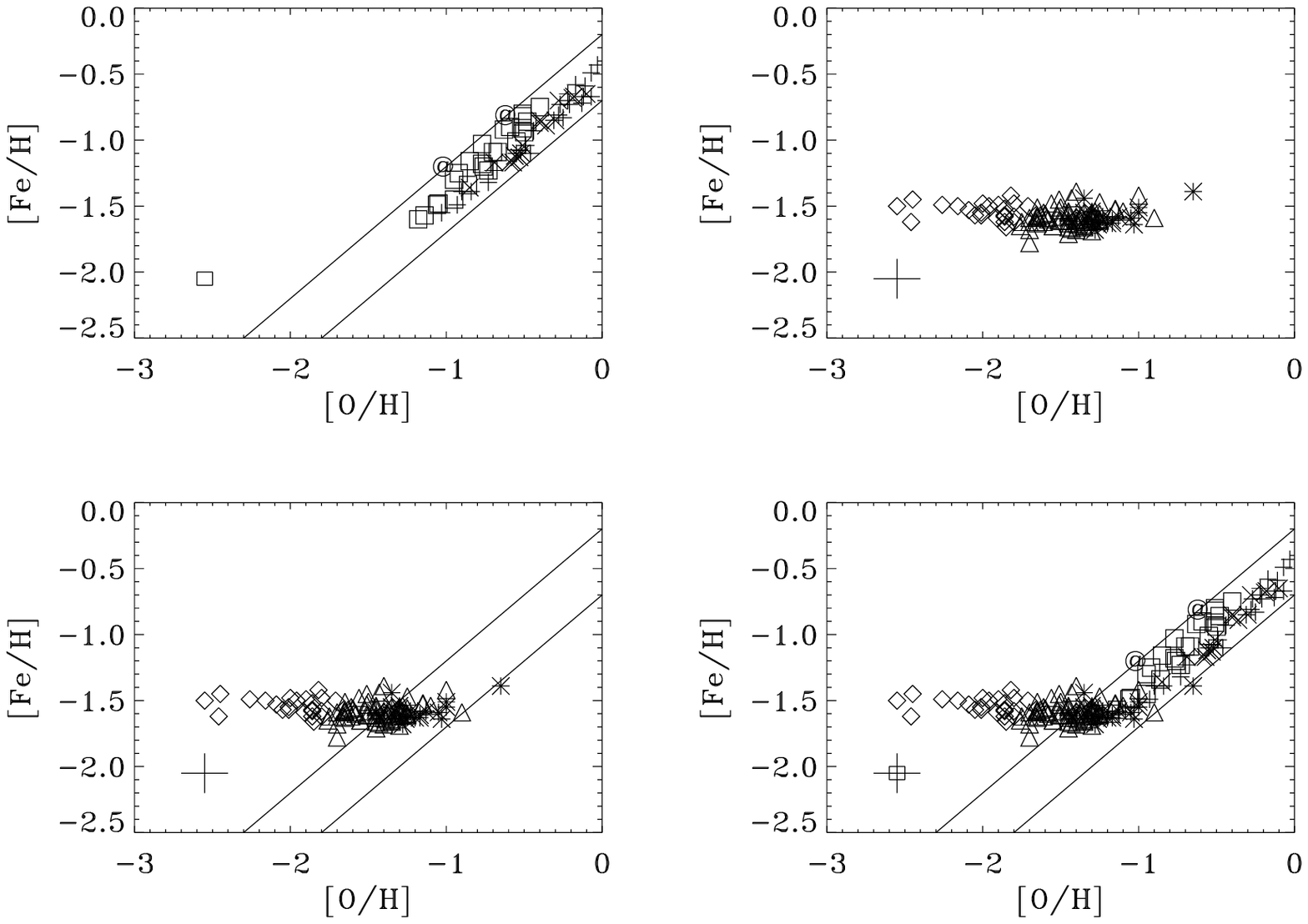}                      
\caption[ddbb]{The selected main sequence, [Fe/H] = [O/H]$-0.45\mp0.25$,
together with the [Fe/H]-[O/H] relation for LH (squares), HH (crosses), KD
(saltires), OL (``at'' symbols) populations, inferred (Caimmi 2013) from
recent
investigations (NS10; Ra12), top left panel; the [Fe/H]-[O/H] relation for
primordial (asterisks), intermediate (triangles), extreme (diamonds) M13
stars, inferred from recent subsampled data (JP12), top right panel; as in top
right panel with the main sequence, [Fe/H] = [O/H]$-0.45\mp0.25$,
superimposed, bottom left panel; as in bottom left panel with LH, HH, KD, OL
data superimposed, bottom right panel.   Typical errors are represented as a
cross (M13) and a square (LH, HH, KD, OL) when appropriate, on the bottom left
of each panel.   For further details refer to the text.}
\label{f:jnrofe4}     
\end{center}       
\end{figure*}

\begin{figure*}[t]  
\begin{center}      
\includegraphics[scale=0.8]{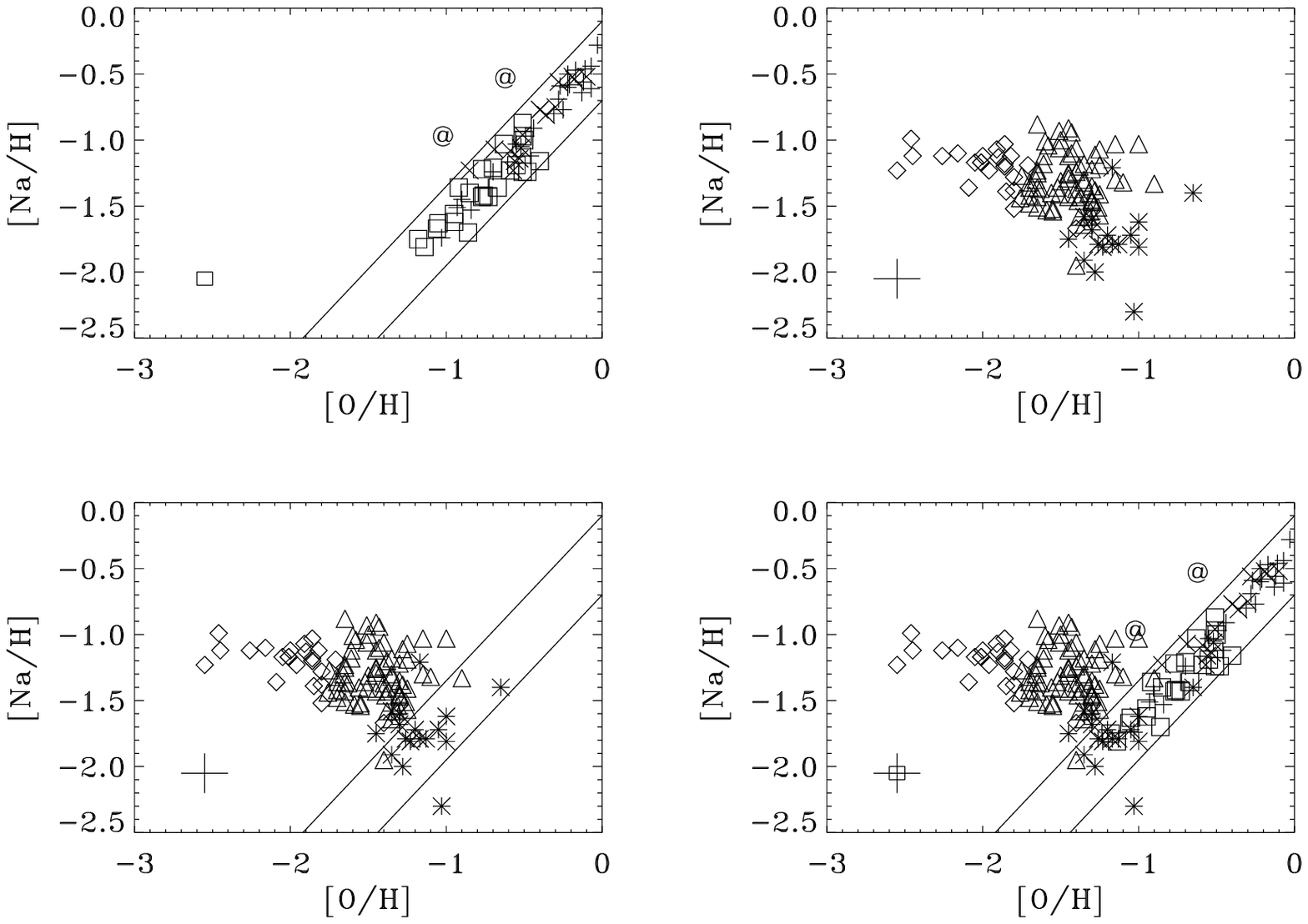}                      
\caption[ddbb]{The selected main sequence, [Na/H] = [O/H]$-0.4\mp0.3$,
together with the [Na/H]-[O/H] relation for LH (squares), HH (crosses), KD
(saltires), OL (``at'' symbols) populations, inferred (Caimmi 2013) from
recent
investigations (NS10; Ra12), top left panel; the [Na/H]-[O/H] relation for
primordial (asterisks), intermediate (triangles), extreme (diamonds) M13
stars, inferred from recent subsampled data (JP12), top right panel; as in top
right panel with the main sequence, [Na/H] = [O/H]$-0.4\mp0.3$,
superimposed, bottom left panel; as in bottom left panel with LH, HH, KD, OL
data superimposed, bottom right panel.   Typical errors are represented as a
cross (M13) and a square (LH, HH, KD, OL) when appropriate, on the bottom left
of each panel.   For further details refer to the text.}
\label{f:jnrona4}     
\end{center}       
\end{figure*}

Aiming to an application to M13, where O, Na, Fe number abundances have
recently been determined for about one hundred stars (JP12), the current
investigation shall be restricted to Q = Na, Fe, keeping in mind it can be
extended to any element for which number abundances are available for a large
star sample.   The main sequences chosen for iron and sodium, [Fe,O] =
[1.00, $-$0.45, 0.50] and [Na,O] = [1.25, $-$0.40, 0.60], are shown in
Figs.\,\ref{f:jnrofe4} and \ref{f:jnrona4}, respectively, top left panels,
together with data from LH (squares), HH (crosses), KD (saltires), OL (``at''
symbols) subsamples, taken from a current attempt (Caimmi 2013).   The
selected slopes are a compromise between related regression lines (Caimmi
2013).

GC stars are currently subsampled into three populations according if the O-Na
anticorrelation is absent, weak, strong, defined as primordial (P),
intermediate (I), extreme (E), respectively (e.g., JP12).   In general, GC
stars are classified as normal, if element abundance is similar to their field
halo counterparts with equal iron abundance, and anomalous if otherwise (e.g.,
C12).   In both cases, no general consensus still exists on a selection
criterion.

With respect to a selected element, Q, let GC P stars be defined as those
lying within the main sequence defined by field halo stars on the
(${\sf O}$\,[O/H]\,[Q/H]) plane, I stars as lying within a parallel
sequence towards lower [O/H], E stars as lying within further parallel
sequences towards lower [O/H].   The above mentioned selection criterion
appears to be natural and well motivated, in that it relates to the main
sequence defined by field halo stars, as shown in Figs.\,\ref{f:jnrofe4} and
\ref{f:jnrona4}, top left panels, for Q = Fe, Na, respectively.   More
specifically, the (${\sf O}$\,[O/H]\,[Q/H]) plane could be divided into an
infinite number of parallel sequences, as:
\begin{lefteqnarray}
\label{eq:OQi}
&& {\rm[Q/H]}=a_{\rm Q}{\rm[O/H]}+b_{\rm Q}+\frac{2i\mp1}2\Delta b_{\rm Q}~~;
\qquad i=0,\mp1,\mp2,...~~;
\end{lefteqnarray}
where $i=0$ labels the main sequence populated by field halo stars, $i<0$ and
$i>0$ label parallel sequences towards larger and lower [O/H], respectively, for
fixed [Q/H].   In this view, a generic GC star can be classified as belonging
to a sequence labelled by an integer, $i$, with regard to a selected element, Q.

\section{Application to M13}\label{M13}

\subsection{Star classification}\label{stcl}

Oxygen, sodium and iron abundance have recently been determined for a sample
$(N=113)$ of red giant branch and AGB stars in M13
(JP12).   Sample stars are divided therein into three classes according to the
following prescriptions: [Na/Fe] $<$ 0.00 - P; [O/Fe] $<$ 0.15 - E;
[Na/Fe] $\ge$ 0.00 and/or [O/Fe] $\ge$ 0.15 - I.   For further details and
exhaustive presentation, an interested reader is addressed to the parent paper
(JP12).   The following subsamples can be extracted from the parent sample: P
$(N=17)$, I $(N=70)$, E $(N=24)$, where both [O/Fe] and [Na/Fe] are known with
the exception of a (different) single star for both O and Na.

The related [Q/H]-[O/H] relation, Q = Fe, Na, is plotted in
Figs.\,\ref{f:jnrofe4}, \ref{f:jnrona4}, respectively, top right panels,
where P, I, E stars are represented as asterisks, triangles, diamonds,
respectively.   The main sequences, [Fe,O] = [1.00, $-$0.45, 0.50], [Na,O] =
[1.25, $-$0.40, 0.60], are superimposed in bottom left panels of 
Figs.\,\ref{f:jnrofe4}, \ref{f:jnrona4}, respectively.   In addition, LH, HH,
KD, OL stars (already shown in top left panels) are superimposed in bottom
right panels of Figs.\,\ref{f:jnrofe4}, \ref{f:jnrona4}, respectively.

\begin{figure*}[t]  
\begin{center}      
\includegraphics[scale=0.8]{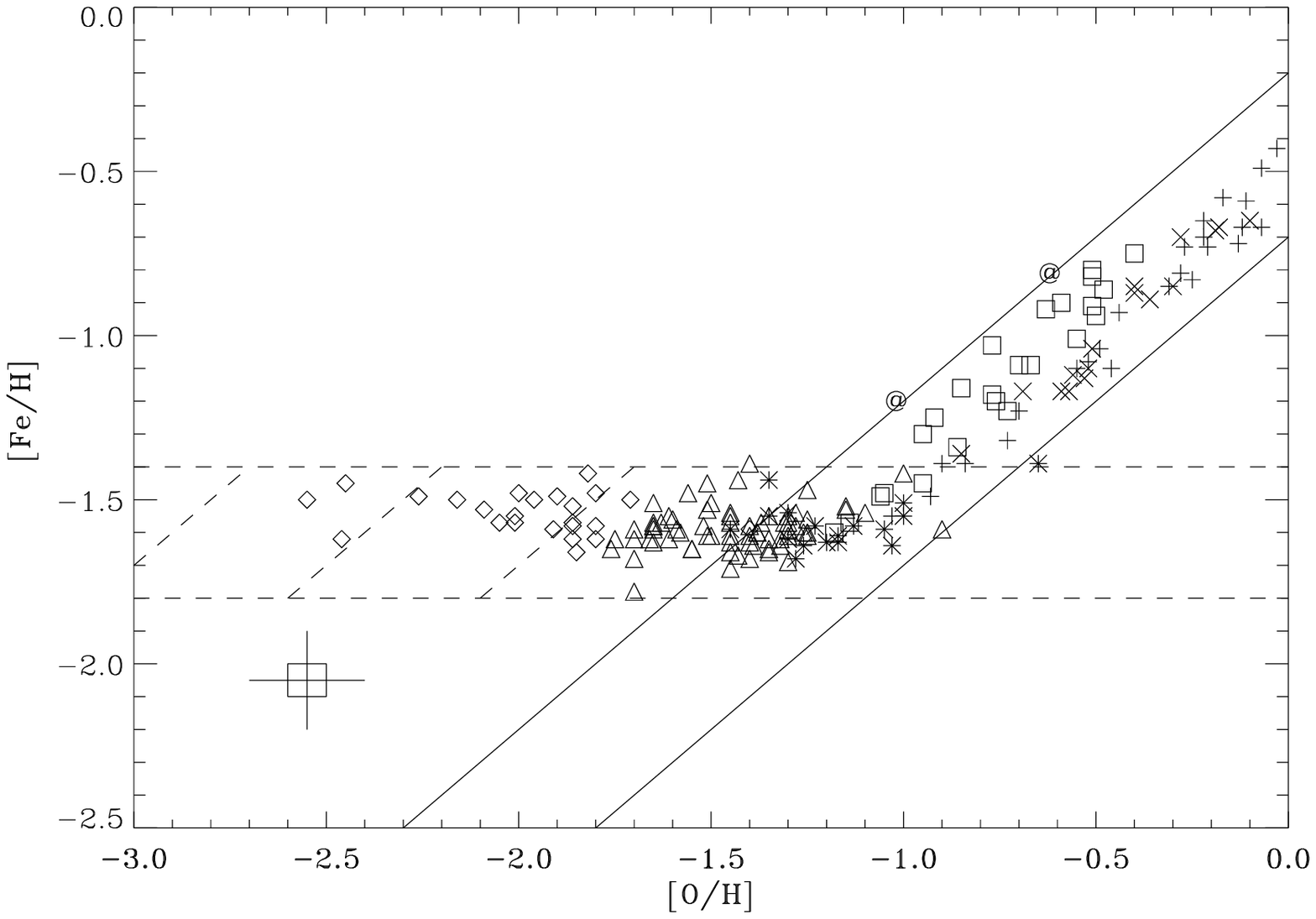}                      
\caption[ddbb]{Zoom of Fig.\,\ref{f:jnrofe4}, bottom right panel, where the
parallel sequences, [Fe/H] = [O/H]$-0.45+(2i\mp1)0.25$,
(class $A_i), i=1,2,3$, and the horizontal branch, [Fe/H] = $-1.6\mp0.2$, are
also plotted (dashed).
M13 sample stars directly shift from the main sequence, [Fe/H] = [O/H]$-0.45
\mp0.25$, (class $A_0$), to the horizontal branch.   For further details
refer to the text.}
\label{f:jnrofe}     
\end{center}       
\end{figure*}

An inspection of Fig.\,\ref{f:jnrofe4} discloses that, within the errors,
P stars and a fraction of I stars lie inside the main sequence,
[Fe/H] = [O/H]$-0.45\mp0.25$, while the remining I stars together with E stars
display along a ``horizontal branch'', [Fe/H] = $-1.6\mp0.2$, which also
encloses the above mentioned stars belonging to the main sequence, as shown in
Fig.\,\ref{f:jnrofe}.
Accordingly, the following star classification can be made with respect to Fe:
normal stars (class A$_0$) as lying within the main sequence, [Fe,O] =
[1.00, $-$0.45, 0.50]; anomalous stars (class A$_i$) as lying within the
parallel sequence, [Fe/O] = [1.00, $-0.45+0.50i, 0.50$].

Let
([O/H],[Fe/H]) be coordinates of a generic sample star on the (${\sf O}$[O/H]
[Fe/H]) plane.   The straight line of unit slope, passing through that point,
has intercept, $b_{\rm Fe}=$ [Fe/H] $-$ [O/H], and the related sequence
is defined by the inequality:
\begin{lefteqnarray}
\label{eq:bfe}
&& -0.7+0.5i\le b_{\rm Fe}<-0.2+0.5i~~;
\end{lefteqnarray}
where $i=-1,1,2,3$, in the case under discussion.   It is worth noticing M13
sample stars directly shift from the main sequence, [Fe/H] = [O/H] $-0.45\mp
0.25$, to the horizontal branch, [Fe/H] = $-1.6\mp0.2$, as shown
in Fig.\,\ref{f:jnrofe}.

\begin{figure*}[t]  
\begin{center}      
\includegraphics[scale=0.8]{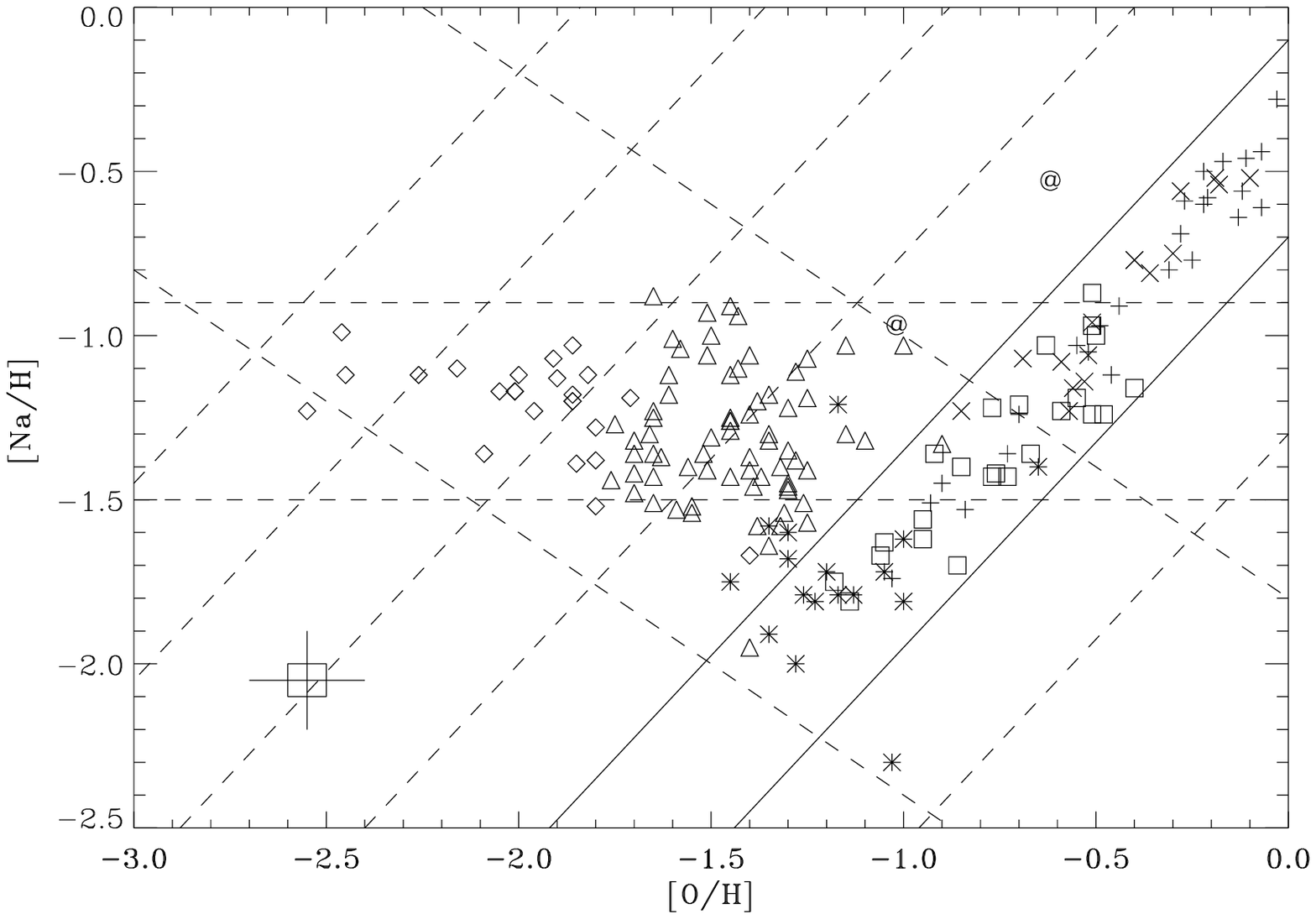}                      
\caption[ddbb]{Zoom of Fig.\,\ref{f:jnrona4}, bottom right panel, where the
parallel sequences, [Na/H] = 1.25\,[O/H]$-0.40+(2i\mp1)0.30$,
(class $A_i$),
$i=-1,1,2,3,4$, the horizontal branch, [Na/H] = $-1.2\,\mp\,0.3$, and the O-Na
anticorrelation, [Na/H] = $-0.8\,[$O/H]$-2.5\mp0.7$, are also plotted (dashed).
M13 sample stars shift from the main sequence, [Na/H] = 1.25\,[O/H]$-0.40\,
\mp\,0.30$, (class $A_0$), to the horizontal branch via the parallel
sequence, [Na/H] = 1.25\,[O/H]\,+\,$0.20\mp0.30$.   For further details
refer to the text.}
\label{f:jnrona}     
\end{center}       
\end{figure*}

An inspection of Fig.\,\ref{f:jnrona4} discloses that, within the errors, a
large fraction of P stars and a few I stars lie inside the main sequence,
[Na/H] = 1.25\,[O/H] $-0.4\mp0.3$, the remining P, I, and almost all E stars
lie inside an inclined band which defines the O-Na anticorrelation,
[Na/H] = $-0.8\,[$O/H]$-2.5\mp0.7$, and ends on a horizontal branch,
[Na/H] = $-1.2\mp0.3$, as shown in Fig.\,\ref{f:jnrona}.
Accordingly, the following star classification can be made with respect to Na:
normal stars (class A$_0$) as lying within the main sequence, [Na,O] =
[1.25, $-$0.40, 0.60]; anomalous stars (class A$_i$) as lying within the
parallel sequence, [Na/O] = [1.25, $-0.40+0.60i, 0.60$].

Let
([O/H],[Na/H]) be coordinates of a generic sample star on the (${\sf O}$[O/H]
[Na/H]) plane.   The straight line of slope, $a_{\rm Na}=1.25$, passing
through that point, has intercept, $b_{\rm Na}=$ [Na/H] $-1.25$ [O/H], and the
related sequence is defined by the inequality:
\begin{lefteqnarray}
\label{eq:bna}
&& -0.7+0.6i\le b_{\rm Na}<-0.1+0.6i~~;
\end{lefteqnarray}
where $i=-1,1,2,3,4$, in the case under discussion.   It is worth noticing M13
sample stars shift from the main sequence, [Na/H] = 1.25\,[O/H] $-0.40\mp
0.30$, to the horizontal branch, [Na/H] = $-1.2\mp0.3$, via the parallel
sequence, [Na/H] = 1.25\,[O/H] $+0.20\mp0.30$, as shown in
Fig.\,\ref{f:jnrona}.   The O-Na anticorrelation, 
[Na/H] = $-$0.8\,[O/H]$-2.5\mp0.7$, is also plotted therein.

An inspection of Figs.\,\ref{f:jnrofe} and \ref{f:jnrona} suggests the
following classification for M13 stars, which can be generalized to any GC
where abundances have been determined for sufficiently large star samples.
Let class N (normal stars), class T (transition stars), class H (horizontal
branch stars), be defined in terms of subclasses as:
\begin{leftsubeqnarray}
\slabel{eq:N}
&& {\rm N}=\sum_{i=-1}^0\sum_{j=-1}^0(A_i,A_j)~~; \\
\slabel{eq:T}
&& {\rm T}=\sum_{i=1}^3(A_i,A_0)+\sum_{j=1}^4(A_0,A_j)~~; \\
\slabel{eq:H}
&& {\rm H}=\sum_{i=1}^3\sum_{j=1}^4(A_i,A_j)~~;
\label{seq:NTH}
\end{leftsubeqnarray}
where $(A_i,A_j)$ relate to Fe and Na classification, respectively, of a
selected sample star and the sum or the double sum extends over the whole set
of possibilities for an assigned class.
Accordingly, a selected sample star is defined by the coordinates,       
$(A_i,A_j)$, $-1\le i\le3$, $-1\le j\le4$, $i$ and $j$ integers, where the  
first and the second place within brackets relate to Fe and Na, respectively.
The whole set of JP12 sample star data used in the current note is reported in
Appendix \ref{a:data}.

The partition of star classes, $(A_i,A_j)$, into populations, P, I, E, as   
defined in the parent paper (JP12), is shown in Table \ref{t:c14m13}.   The 
partition of star classes, N, T, H, as defined by Eq.\,(\ref{seq:NTH}), into
populations, P, I, E, is shown in Table \ref{t:c15m13}.                     
\begin{table}                                                               
\caption{Partition of different M13 star populations according to the parent
paper (JP12), P
(primitive), I (intermediate), E (extreme), into classes with different degree
of anomaly, $A_i$, $i=0,\mp1,\mp2,...$, with regard to Q = Fe, Na, as defined
in the text.   Class $A_{-1}$ is listed as $-A_1$ to save aesthetics.
}
\label{t:c14m13}
\begin{center}
\begin{tabular}{rrrrrr} \hline
\multicolumn{1}{c}{pop:} &
\multicolumn{1}{c}{ } &
\multicolumn{1}{c}{P} &
\multicolumn{1}{c}{I} &
\multicolumn{1}{c}{E} &
\multicolumn{1}{c}{all} \\
\hline
Fe & Na &    &    &    &    \\
\hline
        &        &    &    &    &     \\
$-A_1$ & $ A_0$ &  1 &  0 &  0 &   1 \\
$ A_0$ & $-A_1$ &  1 &  0 &  0 &   1 \\
$ A_0$ & $ A_0$ &  9 &  2 &  0 &  11 \\
$ A_0$ & $ A_1$ &  3 & 25 &  0 &  28 \\
$ A_0$ & $ A_2$ &  0 &  5 &  0 &   5 \\
$ A_1$ & $ A_0$ &  1 &  0 &  0 &   1 \\
$ A_1$ & $ A_1$ &  2 &  7 &  1 &  10 \\
$ A_1$ & $ A_2$ &  0 & 30 &  4 &  34 \\
$ A_1$ & $ A_3$ &  0 &  1 &  3 &   4 \\
$ A_2$ & $ A_2$ &  0 &  0 &  1 &   1 \\
$ A_2$ & $ A_3$ &  0 &  0 & 11 &  11 \\
$ A_2$ & $ A_4$ &  0 &  0 &  1 &   1 \\
$ A_3$ & $ A_4$ &  0 &  0 &  3 &   3 \\
all    &        & 17 & 70 & 24 & 111 \\
\hline                            
\end{tabular}                     
\end{center}                      
\end{table}                       
\begin{table}
\caption{Partition of different M13 star populations according to the parent
paper (JP12), P
(primitive), I (intermediate), E (extreme), into classes N (normal), T
(transition), H (horizontal branch), as defined in
the text.}
\label{t:c15m13}
\begin{center}
\begin{tabular}{lrrrr} \hline
\multicolumn{1}{c}{pop:} &
\multicolumn{1}{c}{P} &
\multicolumn{1}{c}{I} &
\multicolumn{1}{c}{E} &
\multicolumn{1}{c}{all} \\
class &      &    &    &     \\
\hline
    &    &    &    &     \\
N   & 11 &  2 &  0 &  13 \\
T   &  4 & 30 &  0 &  34 \\
H   &  2 & 38 & 24 &  64 \\
all & 17 & 70 & 24 & 111 \\
\hline                            
\end{tabular}                     
\end{center}                      
\end{table}                       
It can be seen class N hosts about two thirds of P stars together with a few
I stars, while class T hosts about one quarter of P stars and slightly less
than one half of I stars.   On the other hand, class H hosts a few P stars,
slightly more than one half of I stars and the whole amount of E stars.

In conclusion, a classification of M13 sample stars as N, T, H, instead of
P, I, E, seems more complete in that it includes both O, Fe, Na, and rigorous
in that it is defined in terms of mean sequences, parallel sequences,
horizontal branches, O-Na anticorrelation, as shown in Figs.\,\ref{f:jnrofe}
and \ref{f:jnrona}.

\subsection{Differential element abundance distribution}\label{dead}

For normalized element mass abundances, $\phi_{\rm Q}=Z_{\rm Q}/
(Z_{\rm Q})_\odot$, the empirical differential abundance distribution reads:
\begin{lefteqnarray}
\label{eq:psie}
&& \psi_{\rm Q}=\log\frac{\Delta N}{N\Delta\phi_{\rm Q}}~~; \\
\label{eq:Dpsie}
&& \Delta^\mp\psi_{\rm Q}=\log\left[1\mp\frac{(\Delta N)^{1/2}}{\Delta N}
\right]~~;
\end{lefteqnarray}
where $\Delta N$ is the number of sample stars binned into
[Q/H]$^\mp=$ [Q/H]$\mp\Delta$[Q/H], $N$ is the sample population, and the
uncertainty on
$\Delta N$, $(\Delta N)^{1/2}$, has been evaluated from Poissonian errors.
In addition, $\log\phi_{\rm Q}=[$Q/H] to a good extent, which implies the
following:
\begin{lefteqnarray}
\label{eq:phic}
&& \phi_{\rm Q}=\frac12\{\exp_{10}{\rm[Q/H]}^++\exp_{10}{\rm[Q/H]}^-\}~~; \\
\label{eq:phia}
&& \Delta^\mp\phi_{\rm Q}=\frac12\{\exp_{10}{\rm[Q/H]}^+-\exp_{10}{\rm[Q/H]}^
-\}~~;
\end{lefteqnarray}
where the bin, $\Delta\phi_{\rm Q}=\Delta^+\phi_{\rm Q}+\Delta^-\phi_{\rm Q}$,
is variable for fixed bin, $\Delta[$Q/H], and vice versa.   For further
details and complete references, an interested reader is addressed to the
parent papers (Caimmi 2011a, 2012a).

The empirical differential abundance distribution, inferred from
Eq.\,(\ref{eq:psie}) with regard to JP12 sample, is listed in Table
\ref{t:c16m13} for O, Na, and in Table \ref{t:c17m13} for Fe, where bins in
[Q/H] are centered on integer decibels and the bin width is
$\Delta$[Q/H] = 1 dex.   More specifically, $-2.6\le[$O/H$]\le-0.7$;
$-2.3\le[$Na/H$]\le-0.9$;
$-1.8\le[$Fe/H$]\le-1.4$.   The lower/upper uncertainty, $\Delta^\mp\psi$, and
the bin semiamplitude, $\Delta^\mp\phi$, may be determined via
Eqs.\,(\ref{eq:Dpsie}) and (\ref{eq:phia}), respectively.
\begin{table}
\caption{M13 empirical, differential
oxygen and sodium abundance distribution
deduced from the JP12 sample
($N=112$).
The error on the generic bin height
has been estimated from the Poissonian error.
The bin width
in [Q/H] $=\log\phi_{\rm Q}$ is $\Delta$[Q/H] $=1$ dex.
See text
for further details.}
\label{t:c16m13}
\begin{center}
\begin{tabular}{llrlr} \hline
\multicolumn{1}{c|}{}
&\multicolumn{2}{c|}{O}
&\multicolumn{2}{c}{Na} \\
\hline\noalign{\smallskip}
\multicolumn{1}{c}{$\phi$} & \multicolumn{1}{c}{$\phantom{0}\psi$} &
\multicolumn{1}{c}{$\Delta N$} &
\multicolumn{1}{c}{$\phantom{0}\psi$} &
\multicolumn{1}{c}{$\Delta N$} \\
\noalign{\smallskip}
\hline\noalign{\smallskip}                                                                                           
2.528552D$-$03 & $+$1.187607D$+$00 & 01  &                   &    \\
3.183258D$-$03 & $+$1.388637D$+$00 & 02  &                   &    \\
4.007485D$-$03 & $ $         $ $   & 00  &                   &    \\
5.045125D$-$03 & $+$8.876073D$-$01 & 01  & $+$8.876073D$-$01 & 01 \\
6.351436D$-$03 & $+$7.876073D$-$01 & 01  &                   & 00 \\
7.995984D$-$03 & $+$9.886373D$-$01 & 02  &                   & 00 \\
1.006635D$-$02 & $+$1.189667D$+$00 & 04  & $+$8.886373D$-$01 & 02 \\
1.267278D$-$02 & $+$1.265759D$+$00 & 06  & $+$4.876073D$-$01 & 01 \\
1.595408D$-$02 & $+$1.232705D$+$00 & 07  & $+$1.086577D$+$00 & 05 \\
2.008500D$-$02 & $+$1.241850D$+$00 & 09  & $+$9.865773D$-$01 & 05 \\
2.528552D$-$02 & $+$1.266789D$+$00 & 12  & $+$1.032705D$+$00 & 07 \\
3.183258D$-$02 & $+$1.166789D$+$00 & 12  & $+$1.166789D$+$00 & 12 \\
4.007485D$-$02 & $+$1.266361D$+$00 & 19  & $+$1.288637D$+$00 & 20 \\
5.045125D$-$02 & $+$1.142880D$+$00 & 18  & $+$1.091727D$+$00 & 16 \\
6.351436D$-$02 & $+$6.906973D$-$01 & 08  & $+$1.018056D$+$00 & 17 \\
7.995984D$-$02 & $+$2.896673D$-$01 & 04  & $+$8.636986D$-$01 & 15 \\
1.006635D$-$01 & $+$1.896673D$-$01 & 04  & $+$4.327053D$-$01 & 07 \\
1.267278D$-$01 & $-$5.123927D$-$01 & 01  & $-$8.966729D$-$02 & 04 \\
1.595408D$-$01 & $ $         $ $   & 00  &                   &    \\
2.008500D$-$01 & $-$7.123927D$-$01 & 01  &                   &    \\
\noalign{\smallskip}      
\hline                                                       
\end{tabular}                                                
\end{center}                                                 
\end{table}                                                  
\begin{table}
\caption{M13 empirical, differential
iron abundance distribution 
deduced from the JP12 sample
($N=113$).
The error on the generic bin height
has been estimated from the Poissonian error.
The bin width
in [Fe/H] $=\log\phi_{\rm Fe}$ is $\Delta$[Fe/H] $=1$ dex.
See text
for further details.}
\label{t:c17m13}
\begin{center}
\begin{tabular}{llr} \hline
\hline\noalign{\smallskip}
\multicolumn{1}{c}{$\phi$} & \multicolumn{1}{c}{$\phantom{0}\psi$} &
\multicolumn{1}{c}{$\Delta N$} \\
\noalign{\smallskip}
\hline\noalign{\smallskip}                                                                                           
1.595408D$-$02 & $+$3.837469D$-$01 & 01 \\
2.008500D$-$02 & $+$1.237989D$+$00 & 09 \\
2.528552D$-$02 & $+$1.996660D$+$00 & 65 \\
3.183258D$-$02 & $+$1.560868D$+$00 & 30 \\
4.007485D$-$02 & $+$8.868369D$-$01 & 08 \\
\noalign{\smallskip}      
\hline                                                       
\end{tabular}                                                
\end{center}                                                 
\end{table}                                                  

\begin{figure*}[t]  
\begin{center}      
\includegraphics[scale=0.8]{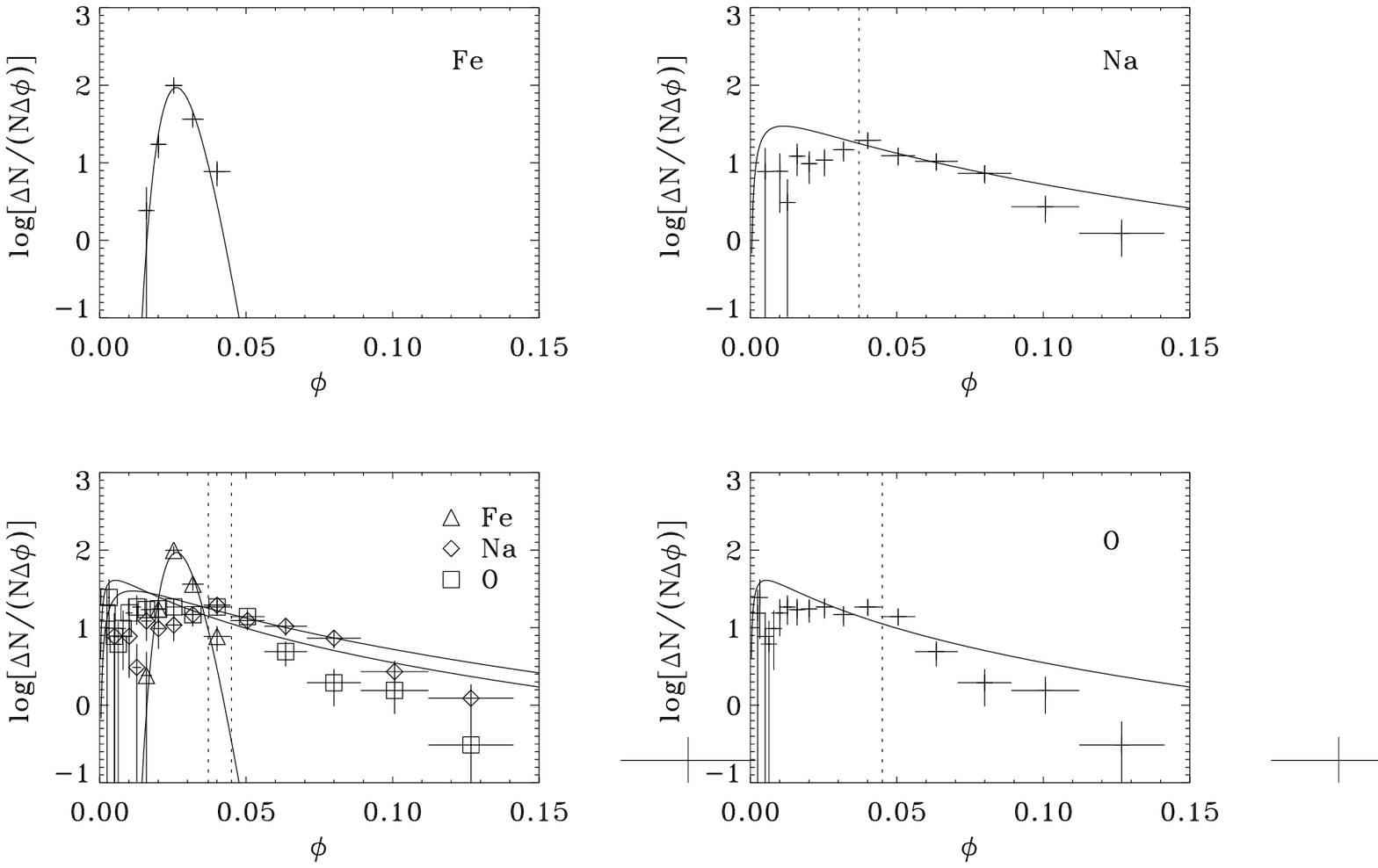}                      
\caption[ddbb]{M13 empirical differential abundance distribution deduced from 
the JP12 sample
related to Fe (top left panel), Na (top right panel), O (bottom right panel),
all together (bottom left panel).   The point out of scale corresponds to
oxygen
abundance within a single star.   The theoretical differential abundance
distribution due to cosmic scatter obeying a Gaussian distribution where
$(x^\ast,\sigma_x)=(<$[Q/H]$>, \sigma_{\rm [Q/H]})$, is also plotted on each
panel for Fe, Na, O, when appropriate.   The dotted vertical lines mark
$\phi_{\rm O}=0.045$ or
[O/H$]=-1.35$ and $\phi_{\rm Na}=0.037$ or [Na/H$]=-1.45$, when appropriate.
For further details refer to the text.}
\label{f:nafeoh4}     
\end{center}       
\end{figure*}

The results are plotted in
Fig.\,\ref{f:nafeoh4} for Fe (top left panel), Na (top right panel), O (bottom
right panel), all together (bottom left panel), respectively.   Also plotted
on each
panel is the theoretical differential abundance distribution due to cosmic
scatter, assumed to obey a Gaussian distribution with parameters inferred
from JP12 sample, as listed in Table \ref{t:mesi}.
\begin{table}                                                               
\caption{Star number, $N$, mean abundance, $\overline{\rm[Q/H]}=\,<$[Q/H]$>$,
rms error,
$\sigma_{\rm [Q/H]}$, Q = O, Na, Fe, inferred from the JP12 sample for star
classes defined above and below in the text.}
\label{t:mesi}
\begin{center}
\begin{tabular}{lrllllll} \hline
 & & & & & & &  \\
\multicolumn{1}{l}{class} &
\multicolumn{1}{r}{$N$} &
\multicolumn{1}{c}{$\overline{\rm[O/H]}$} &
\multicolumn{1}{c}{$\sigma_{\rm [O/H]}$} &
\multicolumn{1}{c}{$\overline{\rm [Na/H]}$} &
\multicolumn{1}{c}{$\sigma_{\rm[Na/H]}$} &
\multicolumn{1}{c}{$\overline{\rm[Fe/H]}$} &
\multicolumn{1}{c}{$\sigma_{\rm [Fe/H]}$} \\
\hline
N      &  13 & $-$1.10 & 0.19 & $-$1.77 & 0.25 & $-$1.59 & 0.07 \\
T      &  34 & $-$1.30 & 0.10 & $-$1.37 & 0.20 & $-$1.60 & 0.06 \\
H      &  64 & $-$1.71 & 0.27 & $-$1.26 & 0.19 & $-$1.56 & 0.07 \\ 
T+H    &  98 & $-$1.57 & 0.30 & $-$1.30 & 0.20 & $-$1.57 & 0.07 \\
N+T+H  & 111 & $-$1.51 & 0.33 & $-$1.35 & 0.26 & $-$1.57 & 0.07 \\
O$_-$  &  76 & $-$1.59 & 0.28 & $ $     &      & $ $     &      \\
O$_+$  &  36 & $-$1.24 & 0.15 & $ $     &      & $ $     &      \\
Na$_-$ &  33 & $ $     &      & $-$1.66 & 0.19 & $ $     &      \\
Na$_+$ &  79 & $ $     &      & $-$1.22 & 0.12 & $ $     &      \\
\hline                            
\end{tabular}                     
\end{center}                      
\end{table}                       
It can be seen the empirical differential abundance distribution is consistent
with its counterpart due to cosmic scatter for Fe, while the contrary holds
for O and Na.   The point out of scale corresponds to oxygen abundance within
a single star.   The theoretical differential abundance distribution due to
cosmic scatter for a selected element, Q, is explicitly expressed in Appendix
\ref{a:cosca}.

Further inspection of Fig.\,\ref{f:nafeoh4} shows the empirical differential
abundance distribution looks quite similar for O and Na in the sense that,
above and below a threshold, 
$\phi_{\rm O}\appleq0.045$ or [O/H] $\le-1.35$ (class O$_-$) and
$\phi_{\rm Na}\appleq0.037$ or [Na/H] $\le-1.45$ (class Na$_-$), respectively,
(i) the trend suddenly changes and (ii) a
linear fit to the data may safely be performed.   Interestingly, N stars
exhibit [O/H] $>-1.35$ for 12/13 of the total and [Na/H] $<-1.45$ for 11/13 of
the total.   Conversely, H stars exhibit [O/H] $\le-1.35$
but [Na/H] $>-1.45$ (class Na$_+$) holds for 55/64 of the total.   Finally, T
stars exhibit [O/H] $>-1.35$ (class O$_+$) and [Na/H] $<-1.45$ for 20/34 and
11/34 of the total,
respectively.   The whole set of results is listed in Table \ref{t:c18m13}.
\begin{table}
\caption{Partition of different M13 star classes, N (normal), T (transition),
H (horizontal branch), into classes, O$_-$, O$_+$, Na$_-$, Na$_+$, as defined
in the text.}
\label{t:c18m13}
\begin{center}
\begin{tabular}{lrrrrr} \hline
\multicolumn{1}{c}{} &
\multicolumn{1}{c}{O$_-$} &
\multicolumn{1}{c}{O$_+$} &
\multicolumn{1}{c}{Na$_-$} &
\multicolumn{1}{c}{Na$_+$} &
\multicolumn{1}{c}{all} \\
\hline
    &    &    &    &    &     \\
N   &  1 & 12 & 11 &  2 &  13 \\
T   & 14 & 20 & 11 & 23 &  34 \\
H   & 64 &  0 &  9 & 55 &  64 \\
all & 79 & 32 & 31 & 80 & 111 \\
\hline                            
\end{tabular}                     
\end{center}                      
\end{table}                       

The upper tail of oxygen distribution, $-0.9\le$ [O/H]$ \le-0.7$, consists of
one star per bin, for a total of two, placed on the lower boundary of the
main sequence, [Fe,O], plotted in Fig.\,\ref{f:jnrofe}, which might be outliers.   
The lower tail of oxygen distribution, $-2.6\le$ [O/H] $\le-2.2$, consists of
no, one, or at most two stars per bin, for a total of five, placed on the
extreme left of the horizontal branch plotted in Fig.\,\ref{f:jnrofe}, which
might be outliers.

\begin{figure*}[t]  
\begin{center}      
\includegraphics[scale=0.8]{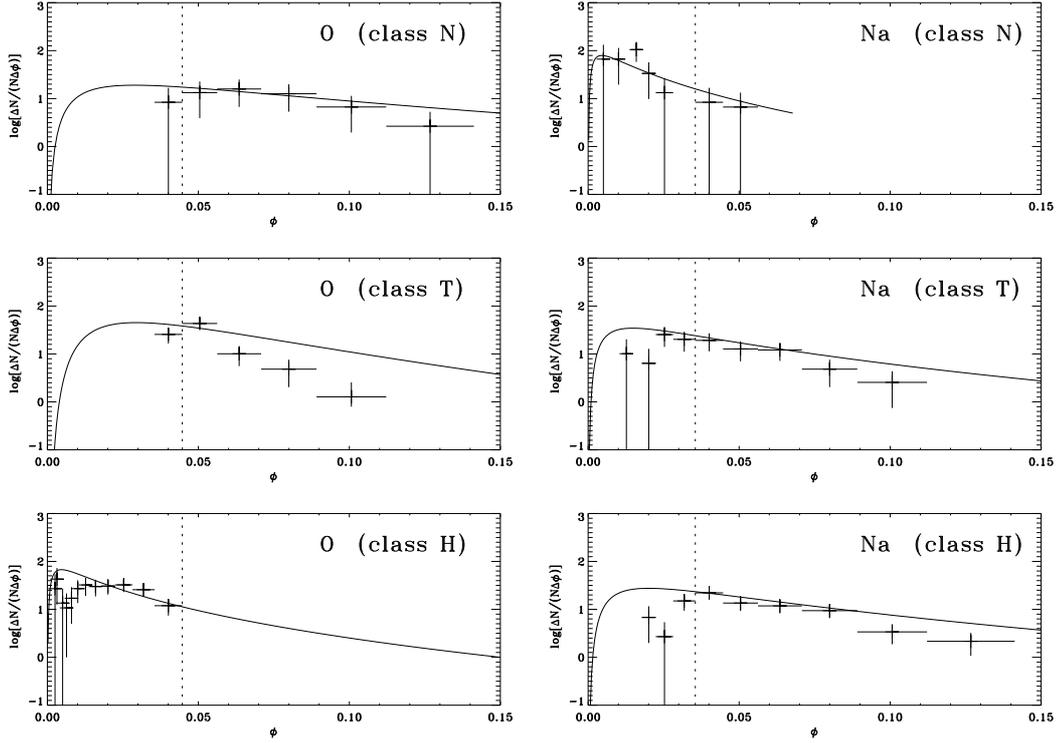}                      
\caption[ddbb]{M13 empirical differential abundance distribution deduced from 
the JP12 sample for stars belonging to class N $(N=13)$, T $(N=34)$, 
H $(N=64)$, related to O ( left panels) and Na ( right panels).
The theoretical differential
distribution due to cosmic scatter obeying a Gaussian distribution where
$(x^\ast,\sigma_x)=(<$[Q/H]$>, \sigma_{\rm [Q/H]})$, is also plotted on each
panel for O, Na, when appropriate.   The dotted vertical lines mark
$\phi_{\rm O}=0.045$ or
[O/H$]=-1.35$ and $\phi_{\rm Na}=0.035$ or [Na/H$]=-1.45$, when appropriate.
For further details refer to the text.}
\label{f:naoh6} 
\end{center}       
\end{figure*}

\begin{figure*}[t]  
\begin{center}      
\includegraphics[scale=0.8]{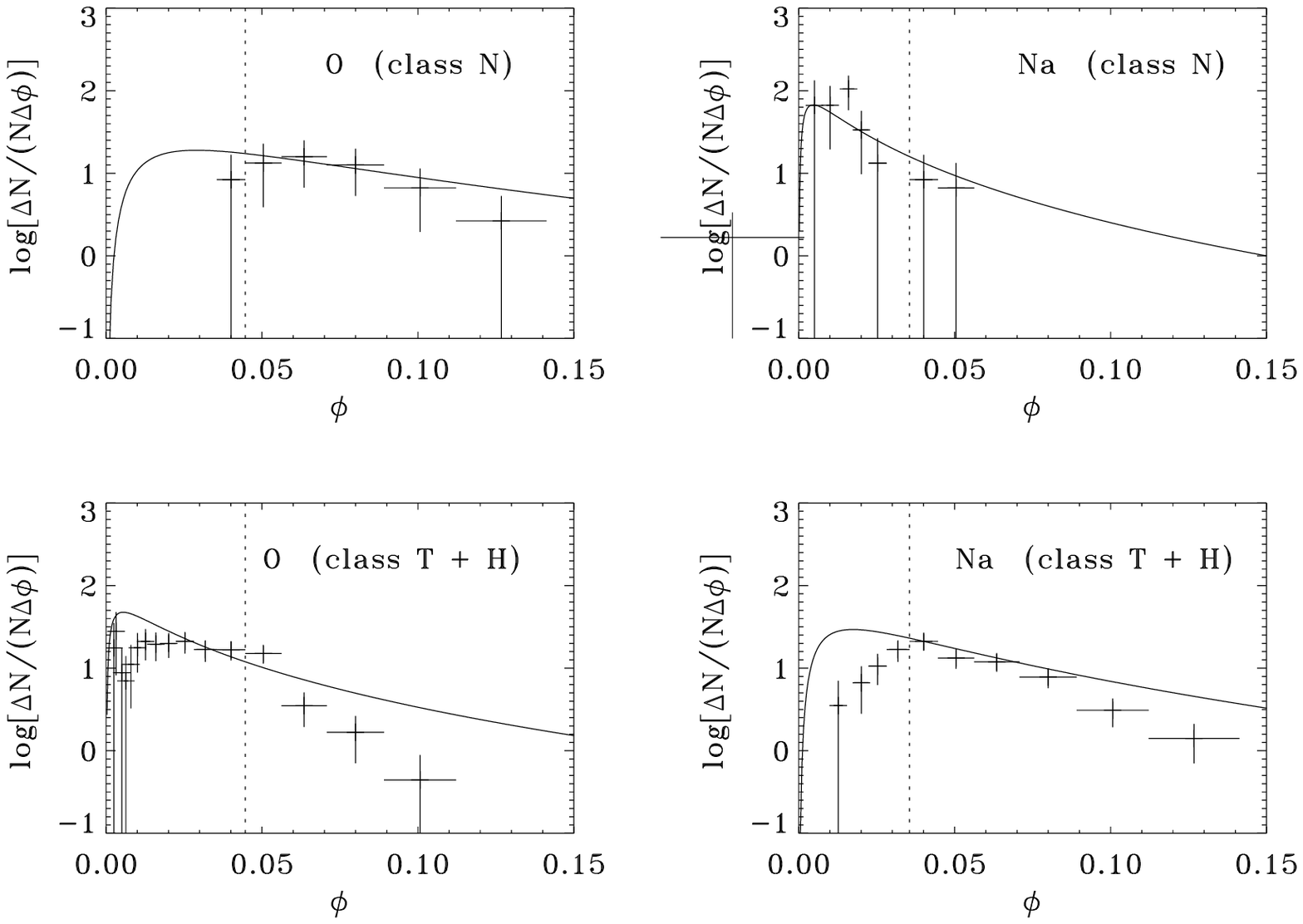}                   
\caption[ddbb]{M13 empirical differential abundance distribution deduced from 
the JP12 sample for stars belonging to class N $(N=13)$,
T+H $(N=98)$, related to O ( left panels) and Na ( right panels).
The point out of scale corresponds to oxygen abundance within a single star.
Other captions as in Fig.\,\ref{f:naoh6}.}
\label{f:naoh4a} 
\end{center}       
\end{figure*}

The empirical differential abundance distribution, deduced from the JP12
sample for stars belonging to class N, T, H, T + H, is listed in Tables 
\ref{t:c16ab3} and \ref{t:c16c13} for O and Na, respectively, where bin width
and uncertainties are taken as in Table \ref{t:c16m13}.
\begin{table}
\caption{M13 empirical, differential
oxygen abundance distribution
deduced from the JP12 sample for stars belonging to class N $(N=13)$, T
$(N=34)$, H $(N=64)$, T + H $(N=98)$.   Errors as in Table \ref{t:c16m13} and
$\delta=\Delta N$ to save space.
See text
for further details.}
\label{t:c16ab3}
\begin{center}
\begin{tabular}{llrlrlrlr} \hline
\multicolumn{1}{c|}{}
&\multicolumn{2}{c|}{class N}
&\multicolumn{2}{c|}{class T}
&\multicolumn{2}{c|}{class H}
&\multicolumn{2}{c}{class T + H} \\
\hline\noalign{\smallskip}
\multicolumn{1}{c}{$\phi$} & 
\multicolumn{1}{c}{$\phantom{0}\psi$} &
\multicolumn{1}{c}{$\delta$} &
\multicolumn{1}{c}{$\phantom{0}\psi$} &
\multicolumn{1}{c}{$\delta$} &
\multicolumn{1}{c}{$\phantom{0}\psi$} &
\multicolumn{1}{c}{$\delta$} &
\multicolumn{1}{c}{$\phantom{0}\psi$} &
\multicolumn{1}{c}{$\delta$} \\
\noalign{\smallskip}
\hline\noalign{\smallskip}                                                                                           
2.53D$-$3 & $ $     $ $  &    & $ $     $ $  &    & $+$1.43D$+$0 &  1 & $+$1.25D$+$0 &  1 \\
3.18D$-$3 & $ $     $ $  &    & $ $     $ $  &    & $+$1.63D$+$0 &  2 & $+$1.45D$+$0 &  2 \\
4.01D$-$3 & $ $     $ $  &    & $ $     $ $  &    & $ $     $ $  &  0 & $ $     $ $  &  0 \\
5.05D$-$3 & $ $     $ $  &    & $ $     $ $  &    & $+$1.13D$+$0 &  1 & $+$9.46D$-$1 &  1 \\
6.35D$-$3 & $ $     $ $  &    & $ $     $ $  &    & $+$1.03D$+$0 &  1 & $+$8.46D$-$1 &  1 \\
8.00D$-$3 & $ $     $ $  &    & $ $     $ $  &    & $+$1.23D$+$0 &  2 & $+$1.05D$+$0 &  2 \\
1.01D$-$2 & $ $     $ $  &    & $ $     $ $  &    & $+$1.43D$+$0 &  4 & $+$1.25D$+$0 &  4 \\
1.27D$-$2 & $ $     $ $  &    & $ $     $ $  &    & $+$1.51D$+$0 &  6 & $+$1.32D$+$0 &  6 \\
1.60D$-$2 & $ $     $ $  &    & $ $     $ $  &    & $+$1.48D$+$0 &  7 & $+$1.29D$+$0 &  7 \\
2.01D$-$2 & $ $     $ $  &    & $ $     $ $  &    & $+$1.48D$+$0 &  9 & $+$1.30D$+$0 &  9 \\
2.53D$-$2 & $ $     $ $  &    & $ $     $ $  &    & $+$1.51D$+$0 & 12 & $+$1.32D$+$0 & 12 \\
3.18D$-$2 & $ $     $ $  &    & $ $     $ $  &    & $+$1.41D$+$0 & 12 & $+$1.22D$+$0 & 12 \\
4.01D$-$2 & $+$9.23D$-$1 &  1 & $+$1.41D$+$0 &  8 & $+$1.08D$+$0 &  7 & $+$1.22D$+$0 & 15 \\
5.05D$-$2 & $+$1.12D$+$0 &  2 & $+$1.64D$+$0 & 17 & $ $     $ $  &    & $+$1.18D$+$0 & 17 \\
6.35D$-$2 & $+$1.20D$+$0 &  3 & $+$1.00D$+$0 &  5 & $ $     $ $  &    & $+$5.45D$-$1 &  5 \\
8.00D$-$2 & $+$1.10D$+$0 &  3 & $+$6.82D$-$1 &  3 & $ $     $ $  &    & $+$2.23D$-$1 &  3 \\
1.01D$-$1 & $+$8.24D$-$1 &  2 & $+$1.05D$-$1 &  1 & $ $     $ $  &    & $-$3.54D$-$1 &  1 \\
1.27D$-$1 & $-$4.23D$-$1 &  1 & $ $     $ $  &    & $ $     $ $  &    & $ $     $ $  &    \\
1.60D$-$1 & $ $     $ $  &  0 & $ $     $ $  &    & $ $     $ $  &    & $ $     $ $  &    \\
2.01D$-$1 & $-$2.23D$-$1 &  1 & $ $     $ $  &    & $ $     $ $  &    & $ $     $ $  &    \\
total     & $ $     $ $  & 13 & $ $     $ $  & 34 & $ $     $ $  & 64 & $ $     $ $  & 98 \\ 
\noalign{\smallskip}      
\hline                                                       
\end{tabular}                                                
\end{center}                                                 
\end{table}                                                  
\begin{table}
\caption{M13 empirical, differential
sodium abundance distribution
deduced from the JP12 sample
for stars belonging to class 
N $(N=13)$, T $(N=34)$, H $(N=64)$ and T+H $(N=98)$.   Errors as in Table 
\ref{t:c16m13} and $\delta=\Delta N$ to save space.
See text
for further details.}
\label{t:c16c13}
\begin{center}
\begin{tabular}{llrlrlrlr} \hline
\multicolumn{1}{c|}{}
&\multicolumn{2}{c|}{class N}
&\multicolumn{2}{c|}{class T}
&\multicolumn{2}{c|}{class H}
&\multicolumn{2}{c}{class T + H} \\
\multicolumn{1}{c}{$\phi$} & 
\multicolumn{1}{c}{$\phantom{0}\psi$} &
\multicolumn{1}{c}{$\delta$} &
\multicolumn{1}{c}{$\phantom{0}\psi$} &
\multicolumn{1}{c}{$\delta$} &
\multicolumn{1}{c}{$\phantom{0}\psi$} &
\multicolumn{1}{c}{$\delta$} &
\multicolumn{1}{c}{$\phantom{0}\psi$} &
\multicolumn{1}{c}{$\delta$} \\
\noalign{\smallskip}
\hline\noalign{\smallskip}                                                                                           
5.05D$-$3 & $+$1.82D$+$0 &  1 & $ $     $ $  &    & $ $     $ $  &    & $ $     $ $   &    \\
6.35D$-$3 & $ $     $ $  &  0 &              &    & $ $     $ $  &    &               &    \\
8.00D$-$3 & $ $     $ $  &  0 &              &    & $ $     $ $  &    &               &    \\
1.01D$-$2 & $+$1.82D$+$0 &  2 & $ $     $ $  &    & $ $     $ $  &    & $ $     $ $   &    \\
1.27D$-$2 & $ $     $ $  &  0 & $+$1.01D$+$0 &  1 & $ $     $ $  &    & $+$5.46D$-$01 &  1 \\
1.60D$-$2 & $+$2.02D$+$0 &  5 & $ $     $ $  &  0 & $ $     $ $  &    & $ $     $ $   &  0 \\
2.01D$-$2 & $+$1.52D$+$0 &  2 & $+$8.05D$-$1 &  1 & $+$8.32D$-$1 &  2 & $+$8.23D$-$01 &  3 \\
2.53D$-$2 & $+$1.12D$+$0 &  1 & $+$1.40D$+$0 &  5 & $+$4.31D$-$1 &  1 & $+$1.02D$+$00 &  6 \\
3.18D$-$2 & $ $     $ $  &  0 & $+$1.30D$+$0 &  5 & $+$1.18D$+$0 &  7 & $+$1.22D$+$00 & 12 \\
4.01D$-$2 & $+$9.23D$-$1 &  1 & $+$1.28D$+$0 &  6 & $+$1.34D$+$0 & 13 & $+$1.32D$+$00 & 19 \\
5.05D$-$2 & $+$8.23D$-$1 &  1 & $+$1.10D$+$0 &  5 & $+$1.13D$+$0 & 10 & $+$1.12D$+$00 & 15 \\
6.35D$-$2 & $ $     $ $  &    & $+$1.08D$+$0 &  6 & $+$1.07D$+$0 & 11 & $+$1.08D$+$00 & 17 \\
8.00D$-$2 & $ $     $ $  &    & $+$6.82D$-$1 &  3 & $+$9.72D$-$1 & 11 & $+$8.92D$-$01 & 14 \\
1.01D$-$1 & $ $     $ $  &    & $+$4.06D$-$1 &  2 & $+$5.30D$-$1 &  5 & $+$4.91D$-$01 &  7 \\
1.27D$-$1 & $ $     $ $  &    & $ $     $ $  &    & $+$3.33D$-$1 &  4 & $+$1.48D$-$01 &  4 \\
total     & $ $     $ $  & 13 & $ $     $ $  & 34 & $ $     $ $  & 64 & $ $     $ $   & 98 \\
\noalign{\smallskip}      
\hline                                                       
\end{tabular}                                                
\end{center}                                                 
\end{table}                                                  
The results are plotted in Figs.\,\ref{f:naoh6} and \ref{f:naoh4a} for O (left
panels) and Na (right panels).
Also plotted on each panel is the theoretical differential abundance
distribution due to the cosmic scatter, assumed to obey a Gaussian
distribution, with parameters inferred from the JP12 sample as listed in Table
\ref{t:mesi}.
It can be seen the empirical differential abundance distribution is consistent
with its counterpart due to cosmic scatter, to an acceptable extent, for class
N (O, Na), T (Na), H (Na), while the contrary holds for class T (O), H (O), T
+ H (O, Na).

\begin{figure*}[t]  
\begin{center}      
\includegraphics[scale=0.8]{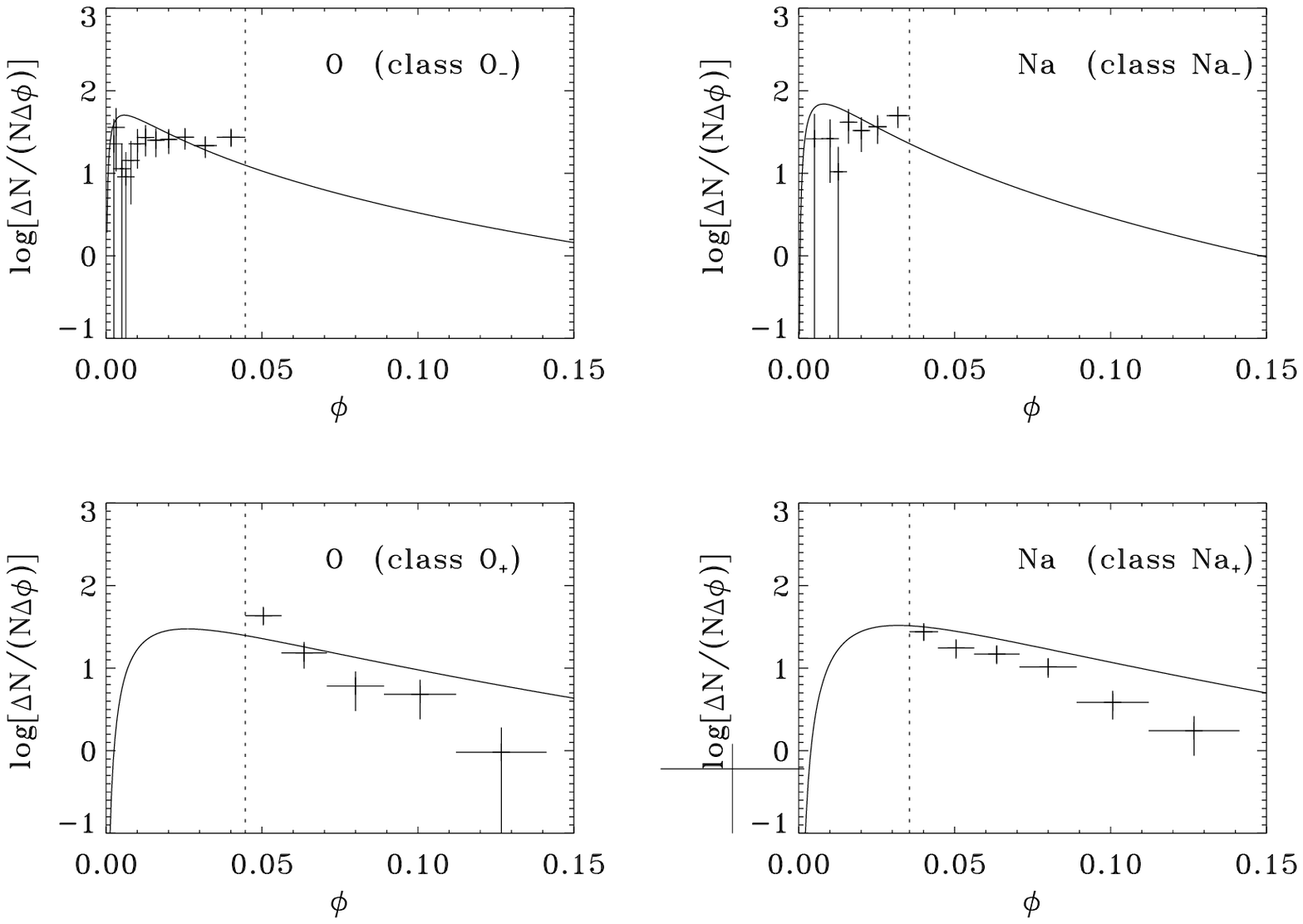}                   
\caption[ddbb]{M13 empirical differential abundance distribution deduced from 
the JP12 sample for stars belonging to class O$_-$ $(N=76)$, O$_+$ $(N=36)$,
Na$_-$ $(N=33)$, Na$_+$ $(N=79)$.
Other captions as in Fig.\,\ref{f:naoh4a}.}
\label{f:naoh4} 
\end{center}       
\end{figure*}

The empirical differential abundance distribution, deduced from the JP12
sample for classes O$_-$, O$_+$, and Na$_-$, Na$_+$, is listed in Table
\ref{t:c16c14} left and right panels, respectively, where bin widths and
uncertainties are taken as in Table \ref{t:c16m13}.
\begin{table}
\caption{M13 empirical, differential
oxygen and sodium abundance distribution
deduced from the JP12 sample for stars belonging to class O$_-$ $(N=36)$,
[O/H] $\le-1.35$ or $\phi\appleq0.040$, O$_+$ $(N=76)$,
[O/H] $>-1.35$ or $\phi\appgeq0.040$, Na$_-$ $(N=33)$,
[Na/H] $\le-1.45$ or $\phi\appleq0.035$, Na$_+$ $(N=79)$,
[Na/H] $>-1.35$ or $\phi\appgeq0.035$.   Errors as in Table \ref{t:c16m13}.
See text for further details.}
\label{t:c16c14}
\begin{center}
\begin{tabular}{llrlr} \hline
\multicolumn{1}{c|}{}
&\multicolumn{2}{c|}{O}
&\multicolumn{2}{c}{Na} \\
\hline\noalign{\smallskip}
\multicolumn{1}{c}{$\phi$} & \multicolumn{1}{c}{$\phantom{0}\psi$} &
\multicolumn{1}{c}{$\Delta N$} &
\multicolumn{1}{c}{$\phantom{0}\psi$} &
\multicolumn{1}{c}{$\Delta N$} \\
\noalign{\smallskip}
\hline\noalign{\smallskip}                                                                                           
2.528552D$-$03 & $+$1.356012D$+$00 & 01  &                   &    \\
3.183258D$-$03 & $+$1.557042D$+$00 & 02  &                   &    \\
4.007485D$-$03 & $ $         $ $   & 00  &                   &    \\
5.045125D$-$03 & $+$1.056012D$+$00 & 01  & $+$1.418311D$+$00 & 01 \\
6.351436D$-$03 & $+$9.560117D$-$01 & 01  &                   & 00 \\
7.995984D$-$03 & $+$1.157042D$+$00 & 02  &                   & 00 \\
1.006635D$-$02 & $+$1.358072D$+$00 & 04  & $+$1.419341D$+$00 & 02 \\
1.267278D$-$02 & $+$1.434163D$+$00 & 06  & $+$1.018311D$+$00 & 01 \\
1.595408D$-$02 & $+$1.401110D$+$00 & 07  & $+$1.617281D$+$00 & 05 \\
2.008500D$-$02 & $+$1.410254D$+$00 & 09  & $+$1.517281D$+$00 & 05 \\
2.528552D$-$02 & $+$1.435193D$+$00 & 12  & $+$1.563409D$+$00 & 07 \\
3.183258D$-$02 & $+$1.335193D$+$00 & 12  & $+$1.697493D$+$00 & 12 \\
4.007485D$-$02 & $+$1.434765D$+$00 & 19  & $+$1.440228D$+$00 & 20 \\

5.045125D$-$02 & $+$1.635795D$+$00 & 18  & $+$1.243318D$+$00 & 16 \\
6.351436D$-$02 & $+$1.183613D$+$00 & 08  & $+$1.169647D$+$00 & 17 \\
7.995984D$-$02 & $+$7.825828D$-$01 & 04  & $+$1.015289D$+$00 & 15 \\
1.006635D$-$01 & $+$6.825828D$-$01 & 04  & $+$5.842963D$-$01 & 07 \\
1.267278D$-$01 & $-$1.947718D$-$02 & 01  & $+$2.412582D$-$01 & 04 \\
1.595408D$-$01 & $ $         $ $   & 00  &                   &    \\
2.008500D$-$01 & $-$2.194772D$-$01 & 01  &                   &    \\
\noalign{\smallskip}      
\hline                                                       
\end{tabular}                                                
\end{center}                                                 
\end{table}                                                  
The results are plotted in Fig.\,\ref{f:naoh4} together with related
theoretical differential abundance distributions due to cosmic scatter,
assumed to obey a Gaussian distribution, with parameters inferred from the
JP12 sample as listed in Table \ref{t:mesi}.   It can be seen the empirical
differential abundance distribution is not consistent with its counterpart due
to cosmic scatter.

The empirical differential oxygen and sodium abundance distribution, plotted
in Fig.\,\ref{f:nafeoh4}, is characterized by the presence
of two stages exhibiting a nearly linear trend, which shall be named
as AF, CE, for increasing element abundance.   Linear fits to each stage,
$\psi=a\phi+b$, are performed using the bisector regression (Isobe et al.
1990; Caimmi 2011b, 2012b),
leaving aside bins related to single stars.   The regression line slope and
intercept estimators and related dispersion estimators are listed in Table
\ref{t:abs} for each stage of the O and Na empirical distributions plotted in
Fig.\,\ref{f:nafeoh4}.
\begin{table}
\caption{Regression line slope and intercept
estimators, $\hat{a}$ and $\hat{b}$, and
related dispersion estimators, $\hat{\sigma}_
{\hat{a}}$, and $\hat{\sigma}_{\hat{b}}$,
for regression models applied to the empirical differential
oxygen and sodium abundance distribution plotted in Fig.\,\ref{f:nafeoh4},
bottom and top right panels, respectively.
The method has dealt with
each stage (XX) separately.   Data points on the
boundary between AF and CE stages are used
for determining regression lines within both of them.   Bins containing a
single star are not considered in the regression.}
\label{t:abs}
\begin{center}
\begin{tabular}{lllllc} \hline
\multicolumn{1}{l}{XX} &
\multicolumn{1}{c}{$\hat{a}$} &
\multicolumn{1}{c}{$\hat{\sigma}_{\hat{a}}$} &
\multicolumn{1}{c}{$\hat{b}$} &
\multicolumn{1}{c}{$\hat{\sigma}_{\hat{b}}$} &
\multicolumn{1}{c}{class} \\ 
\hline
   &                 &              &                 &              &        \\
AF & $-$1.4520 E$+$0 & 1.6466 E$+$0 & $ $1.2600 E$+$0 & 5.4019 E$-$2 & O$_-$  \\
CE & $-$2.0886 E$+$1 & 4.5526 E$+$0 & $ $2.1346 E$+$0 & 3.0319 E$-$1 & O$_+$  \\
AF & $+$1.6409 E$+$1 & 4.9000 E$+$0 & $ $7.2001 E$-$1 & 8.3895 E$-$2 & Na$_-$ \\
CE & $-$1.3443 E$+$1 & 1.1448 E$+$0 & $ $1.8272 E$+$0 & 7.0398 E$-$2 & Na$_+$ \\
\hline       
\end{tabular}
\end{center} 
\end{table}  

\begin{figure*}[t]  
\begin{center}      
\includegraphics[scale=0.8]{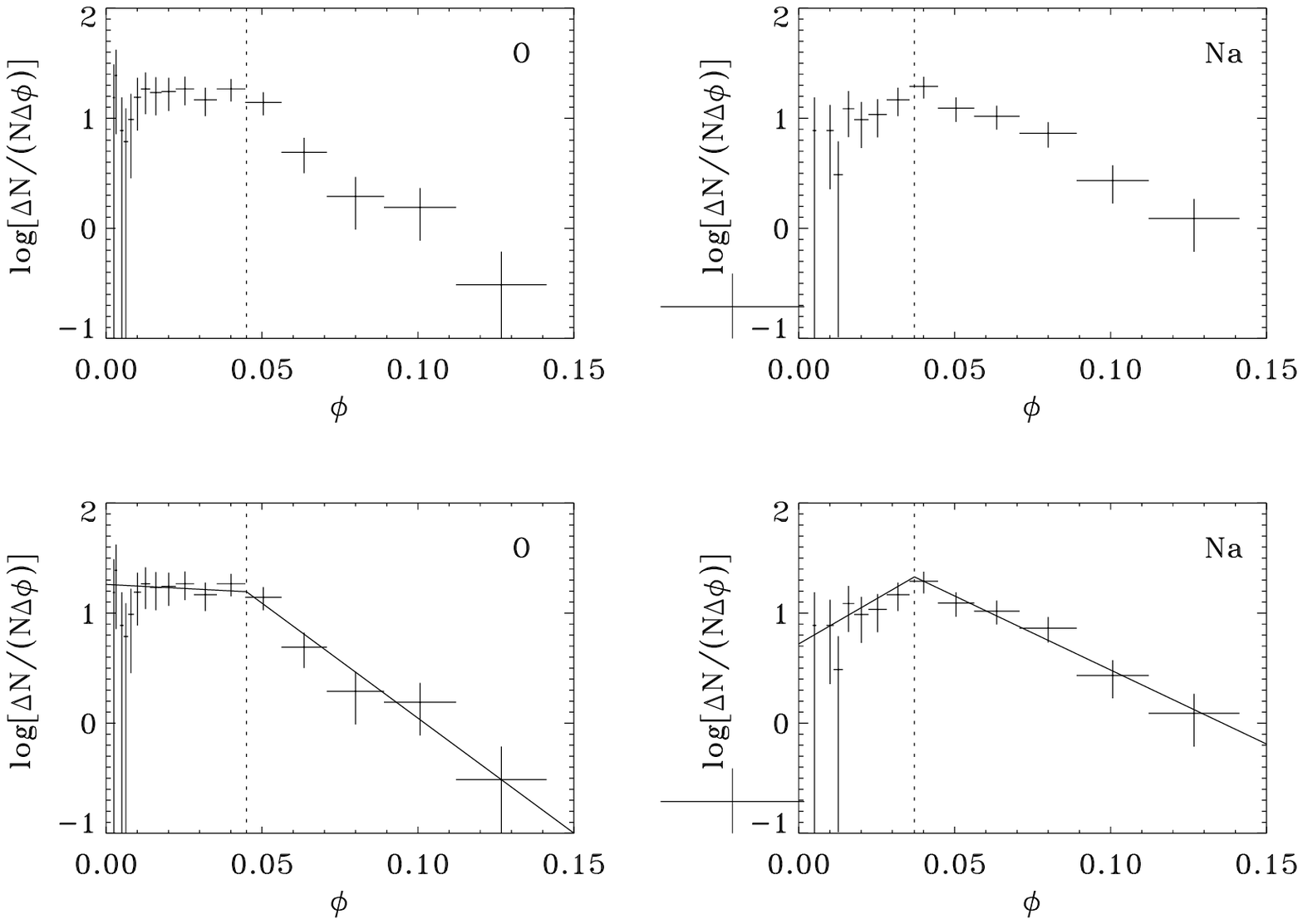}                      
\caption[ddbb]{Regression lines to the empirical
differential oxygen (top left) and sodium (top right) abundance distributions
listed in Table \ref{t:c16m13}
with regard to stages
(from the left to the right) AF and CE.
Bins containing a single
star are not used for the regression.  The abscissa of the 
intersection point of regression lines is marked by a dashed vertical line.
Other captions as in
Fig.\,\ref{f:nafeoh4}.   For further details
refer to the text.}
\label{f:jp12onah4}     
\end{center}       
\end{figure*}

Interesting features shown in Table \ref{t:abs} read (i) the slope of oxygen
distribution above the threshold (CE) and sodium distribution below the
threshold (AF)
are equal and opposite within the errors and (ii) the intercept of oxygen and
sodium distribution above the threshold (CE) are equal within the errors.
The regression lines are represented in Fig.\,\ref{f:jp12onah4} for each
stage, according to slope and intercept values listed in Table \ref{t:abs}.
To ensure continuity, AF and CE stages are bounded by the intersections of
related regression lines.   
The results are listed in Table \ref{t:figs}, where OO denotes element
abundance ranges without data, $0\le\phi\le\phi_i$ or $\phi\ge\phi_f$,
and $\phi_i$ and $\phi_f$ are the minimum and maximum
element abundance, respectively, within sample stars.
\begin{table}
\caption{Transition points between adjacent
stages, as determined from the intersection
of related regression lines, for the empirical differential oxygen and sodium
abundance distribution plotted in
Fig.\,\ref{f:jp12onah4}, top left and right panels, respectively.}
\label{t:figs}
\begin{center}
\begin{tabular}{lllc} \hline
\multicolumn{1}{l}{transition} &
\multicolumn{1}{c}{$\phi$} &
\multicolumn{1}{c}{$\psi$} &
\multicolumn{1}{c}{Q}  \\
\hline
      &              &                 &    \\
OO-AF & 2.5286 E$-$3 & $+$1.2563 E$-$0 & O  \\ 
AF-CE & 4.5002 E$-$2 & $+$1.1947 E$-$0 & O  \\ 
CE-OO & 2.5286 E$-$1 & $-$3.1465 E$-$0 & O  \\ 
OO-AF & 4.4668 E$-$3 & $+$7.9296 E$-$1 & Na \\
AF-CE & 3.7088 E$-$2 & $+$1.3286 E$-$0 & Na \\ 
CE-OO & 1.4125 E$-$1 & $-$7.1773 E$-$2 & Na \\ 
\hline       
\end{tabular}
\end{center} 
\end{table}  
Accordingly, a vertical line instead of a regression line is considered for
the intersection points related to OO-AF and CE-OO transitions.

\section{Discussion} \label{disc}

The GC star classification, defined in Section \ref{GCsc}, is rigorous and
well motivated in that it relates to the main sequence defined by field halo
stars, as shown in Figs.\,\ref{f:jnrofe} and \ref{f:jnrona} in the
case under consideration of M13.   The validity of the proposed classification
may be tested in a twofold manner, namely by extending
Figs.\,\ref{f:jnrofe} and \ref{f:jnrona} to (i) elements other than Na, Fe,
and (ii) GCs other than
M13.   In any case, the expected trend implies part of sample stars within
the above mentioned main sequence, part on a horizontal branch
(Fig.\,\ref{f:jnrofe}) or, in addition, on one or more parallel sequences
towards lower oxygen abundance (Fig.\,\ref{f:jnrona}).   For an assigned
element, different GCs are expected to depart from the main sequence at
different points, according to their primordial abundance in that element.

Caution is needed in the interpretation of quantitative results, owing to
currently large abundance mesaurement errors for GC stars (e.g., Renzini
2013).   Let the trend exhibited by element abundance remain slightly affected
even if the uncertainty on single stars is high, typically $\mp1.5$ dex for
M13
(JP12).   Under this assumption, valuable informations can be extracted from
the empirical differential element abundance distribution, as shown in
Fig.\,\ref{f:jp12onah4} for M13, where a threshold is clearly exhibited at
[O/H] $\approx-1.35$ and [Na/H] $\approx-1.45$, respectively.   More
specifically, different slopes in different regions could reflect the
occurrence of different physical processes during the evolution.

An inspection of Table \ref{t:c18m13} shows the whole amount of H stars is
depleted in oxygen, [O/H] $\le-1.35$ (class O$_-$), while 55/64 of the total
are enriched in sodium, [Na/H] $>-1.45$ (class Na$_+$), which might be
interpreted in a twofold manner:
either sodium enrichment is not the sole channel of oxygen depletion, or
sodium is, in turn, partially depleted.   In fact, the empirical differential
sodium abundance distribution for classes N, T, H, is consistent, to an
acceptable extent, with its theoretical counterpart due to intrinsic scatter,
leaving
aside a few sodium-deficient stars, as plotted in Fig.\,\ref{f:naoh6}.   With
regard to oxygen, a similar trend is exhibited for class N, while the contrary
holds for class T, H.   The above results are consistent with little or no
chemical evolution for oxygen within class N and sodium within class N, T, H.
On the other hand, oxygen chemical evolution did take place, even if
partially, within class T, H, and, in any case, passing from O$_+$ to O$_-$
class and from Na$_-$ to Na$_+$ class.

To this respect, the simplest assumption is a constant yield for
both oxygen and sodium, in the light of simple MCBR (multistage closed box +
reservoir) models of chemical
evolution (Caimmi 2011a, 2012a) where gas inflow and outflow are allowed.
In the special case of a linear fit to the empirical differential
abundance distribution:
\begin{equation}
\label{eq:psi}
\psi_{\rm Q}=a_{\rm Q}\phi_{\rm Q}+b_{\rm Q}~~;
\end{equation}
the slope has the explicit expression:
\begin{equation}
\label{eq:aQ}
a_{\rm Q}=-\frac1{\ln10}\frac{(Z_{\rm Q})_\odot}{\hat{p}_{\rm Q}}(1+\kappa)~~;
\end{equation}
where Q = O, Na, and $\kappa$ is the flow parameter, positive for outflow and
negative for inflow.   For further details refer to the parent papers (Caimmi
2011a, 2012a).

The oxygen-to-sodium yield ratio, inferred from Eq.\,(\ref{eq:aQ}), reads:
\begin{equation}
\label{eq:yONa}
\frac{\hat{p}_{\rm O}}{\hat{p}_{\rm Na}}=\frac{a_{\rm Na}}{a_{\rm O}}
\frac{(Z_{\rm O})_\odot}{(Z_{\rm Na})_\odot}~~;
\end{equation}
where the slope ratio, $a_{\rm Na}/a_{\rm O}$, has necessarily to be negative
for O depletion and Na enrichment, which implies an opposite sign for slopes
related to classes O$_-$, Na$_+$, and O$_+$, Na$_-$, listed in Table
\ref{t:abs}.   Accordingly, the slope related to class O$_-$, within the
errors, restricts to $0<a_{\rm O}\le0.1946$.   Then $a_{\rm Na}/a_{\rm O}=-1$
for stars belonging to classes O$_+$ and Na$_-$, within the errors, while
$a_{\rm Na}/a_{\rm O}\appleq-69$ for stars belonging to classes O$_-$ and
Na$_+$.

The substitution of the above mentioned values into Eq.\,(\ref{eq:yONa})
yields:
\begin{equation}
\label{eq:yOpNam}
\frac{\hat{p}_{\rm O}}{\hat{p}_{\rm Na}}=-\frac{(Z_{\rm O})_\odot}
{(Z_{\rm Na})_\odot}~~;
\end{equation}
in the former alternative and
\begin{equation}
\label{eq:yOmNap}
\frac{\hat{p}_{\rm O}}{\hat{p}_{\rm Na}}\appleq-69\frac{(Z_{\rm O})_\odot}
{(Z_{\rm Na})_\odot}~~;
\end{equation}
in the latter alternative.   According to Eq.\,(\ref{eq:yOpNam}), O depletion
can be entirely turned into Na enrichment during the early evolution ([O/H$]
\ge-1.35$, [Na/H$]\le-1.45$).   Conversely, according to Eq.\,(\ref{eq:yOmNap}),
O depletion appears to be mainly turned into Q other than Na enrichment during
the late evolution ([O/H$]<-1.35$, [Na/H$]>-1.45$).

As an exercise, let sodium chemical evolution in M13 be considered within the
framework of simple MCBR models in the linear limit (Caimmi 2011a, 2012a).
Sodium may be conceived as a primary element provided it is mainly synthesised
from oxygen which is, in turn, a primary element.   Accordingly, the following
relations hold together with Eqs.\,(\ref{eq:psi})-(\ref{eq:yONa}):
\begin{lefteqnarray}
\label{eq:mufi}
&& \frac{\mu_f}{\mu_i}=\exp_{10}\left\{a_{\rm Q}\left[(\phi_{\rm Q})_f-
(\phi_{\rm Q})_i\right]\right\}~~; \\
\label{eq:sfi}
&& s_f-s_i=\frac{(Z_{\rm Q})_\odot}{\hat{p}_{\rm Q}}\mu_i
\left[(\phi_{\rm Q})_f-(\phi_{\rm Q})_i\right]~~; \\
\label{eq:Dfi}
&& D_f-D_i=\kappa(s_f-s_i)~~; \\
\label{eq:zQ}
&& \zeta_{\rm Q}=1-\frac{A_{\rm Q}\hat{p}_{\rm Q}}\kappa~~;
\end{lefteqnarray}
where $\mu$ is the gas mass fraction, $s$ the long-lived star mass fraction,
$D$ the flowing gas mass fraction, $\zeta$ the ratio of Q abundance within the
flowing gas to Q abundance within the pre existing gas, $A_{\rm Q}$ a
coefficient which may safely be expressed as $A_{\rm Q}=2(Z_{\rm O})_\odot/
(Z_{\rm Q})_\odot$, Q = Na in the case under discussion and the indexes, i, f,
denote initial and final values, respectively, with regard to
the stage considered.   The free parameter is chosen to be the true sodium
yield, $\hat{p}_{\rm Na}$.

The cut parameter,
$\zeta_{\rm Na}$, the flow parameter, $\kappa$, the active (i.e. available
for star formation) gas mass fraction,
$\mu_f$, the long-lived star mass fraction, $s_f$, the flowing gas mass
fraction, $D_f$, determined in the linear limit for a true sodium yield in
solar sodium abundance units, $\hat{p}_{\rm Na}/(Z_{\rm Na})_\odot=0.5, 1.0,
2.0$, are listed in Table \ref{t:solin}.
\begin{table}
\caption{Values of the cut parameter, $\zeta_{\rm Na}$,
the flow parameter, $\kappa$, the active gas mass fraction,
$\mu_f$, the long-lived star mass fraction, $s_f$, the flowing gas mass
fraction, $D_f$, for different values of the true sodium yield in 
solar sodium abundance units, $\hat{p}_{\rm Na}/(Z_{\rm Na})_\odot$, in the
linear limit.   For each case, upper lines correspond to Na$_-$ class ([Na/H]
$\le-1.45$; AF stage) and lower lines to Na$_+$ class ([Na/H]$\,>-1.45$; CE
stage).   To save space, $\hat{p}_{\rm Na}$ stands for
$\hat{p}_{\rm Na}/(Z_{\rm Na})_\odot$, XX denotes the stage and cls the class.
For further details refer to the text.}
\label{t:solin}
\begin{center}
\begin{tabular}{llllllll} \hline
\multicolumn{1}{c}{$\hat{p}_{\rm Na}$} &
\multicolumn{1}{c}{$\zeta_{\rm Na}$} &
\multicolumn{1}{c}{$\kappa$} &
\multicolumn{1}{c}{$\mu_f$} &
\multicolumn{1}{c}{$s_f$} &
\multicolumn{1}{c}{$D_f$} &
\multicolumn{1}{c}{XX} &
\multicolumn{1}{l}{cls} \\
\hline
    &             &                &             &             &                &    &        \\
0.5 & 1.0003E$-$0 & $-$1.9893E$+$1 & 3.4302E$-$0 & 1.2863E$-$1 & $-$2.5588E$-$0 & AF & Na$_-$ \\
    & 9.9960E$-$1 & $+$1.4477E$+$1 & 1.3648E$-$1 & 3.4144E$-$1 & $+$5.2208E$-$1 & CE & Na$_+$ \\
    &             &                &             &             &                &    &        \\
1.0 & 1.0003E$-$0 & $-$3.8786E$+$1 & 3.4302E$-$0 & 6.4315E$-$2 & $-$2.4945E$-$0 & AF & Na$_-$ \\
    & 9.9961E$-$1 & $+$2.9954E$+$1 & 1.3648E$-$1 & 1.7072E$-$1 & $+$6.9280E$-$1 & CE & Na$_+$ \\
    &             &                &             &             &                &    &        \\
2.0 & 1.0003E$-$0 & $-$7.6571E$+$1 & 3.4302E$-$0 & 3.2157E$-$2 & $-$2.4623E$-$0 & AF & Na$_-$ \\
    & 9.9962E$-$1 & $+$6.0907E$+$1 & 1.3648E$-$1 & 8.5361E$-$2 & $+$7.7816E$-$1 & CE & Na$_+$ \\
\hline
\end{tabular}                     
\end{center}                      
\end{table}                       
The parameters, $\kappa$, $D_f$, are negative for inflow and positive for
outflow, respectively.   For each case, upper lines correspond to Na$_-$ class
([Na/H] $\le-1.45$; AF stage) and lower lines to Na$_+$ class ([Na/H]
$>-1.45$; CE stage).

Solar abundances are inferred from recent data (Asplund et al.
2009) as $(Z_{\rm Na})_\odot=2.950\,10^{-5}$ and
$(Z_{\rm O})_\odot=5.786\,10^{-3}$ (Caimmi 2013).
The initial values at the AF stage are chosen as $\mu_i=1$;
$s_i=0$; $D_i=0$; accordingly, $\mu_f$ is independent of
$\hat{p}_{\rm Na}/(Z_{\rm Na})_\odot$ via Eq.\,(\ref{eq:mufi}) and the ratio,
$(s_f)_{\rm CE}/(s_f)_{\rm AF}$, is independent of
$\hat{p}_{\rm Na}/(Z_{\rm Na})_\odot$ via Eq.\,(\ref{eq:sfi}).   If
$(s_f)_{\rm CE}$ and $(\mu_f)_{\rm AF}+(s_f)_{\rm AF}$ relate to the current
and the primordial M13 mass, respectively, an inspection of Table
\ref{t:solin} shows the primordial to current mass ratio amounts to about
10, 20, 40, for
$\hat{p}_{\rm Na}/(Z_{\rm Na})_\odot=0.5, 1.0, 2.0$, respectively, according
to Eqs.\,(\ref{eq:mufi})-(\ref{eq:sfi}).   Then the M13 primordial mass can be
inferred, in the framework of the model, from the knowledge of the true sodium
yield.

\section{Conclusion}
\label{conc}

The main results of the current paper may be summarized as follows.
\begin{description}
\item[(1)\hspace{2.0mm}]
Basing on the ``mean sequence'' defined by halo and thick disk low-metallicity
([Fe/H] $<-0.6$) stars on the
(${\sf O}$[O/H][Q/H]) plane (Caimmi 2013), a natural and well motivated
selection criterion has been defined for classifying GC stars with respect to
a selected element, Q, and to oxygen, O.
\item[(2)\hspace{2.0mm}]
An application has been performed to M13 using a star sample $(N=112)$ for
which O, Na, Fe abundance is available (JP12): previously classified
primordial stars are found to lie (leaving aside two exceptions) within the
above mentioned main sequence (class N); previously classified intermediate
stars are found to lie (in comparable amount) within both a parallel main
sequence (towards lower [O/H]; class T) and a ``horizontal branch'' (class H);
previously classified extreme stars are found to lie within the above
mentioned horizontal branch; as shown in Table \ref{t:c15m13} and in
Figs.\,\ref{f:jnrofe}, \ref{f:jnrona}.
\item[(3)\hspace{2.0mm}]
Both O and Na empirical differential abundance distributions have been
determined for each class and the whole sample (with the addition of Fe in the
last case) and compared with their theoretical counterparts due to cosmic
scatter obeying a Gaussian distribution whose parameters were inferred from
related subsamples.   For the whole sample, an acceptable fit has been found
only for Fe, as shown in Fig.\,\ref{f:nafeoh4}, which has been extended to
class N for both O and Na and to class T and H for Na with the exception of a
few sodium-deficient stars, as shown in Fig.\,\ref{f:naoh6}.   No fit occurs
for both T + H class and for stars where [O/H]$<-1.35$ (class O$_-$), [O/H]$
\ge-1.35$ (class O$_+$), [Na/H]$\le-1.45$ (class Na$_-$), [Na/H]$>-1.45$
(class Na$_+$), as shown in Figs.\,\ref{f:naoh4a}, \ref{f:naoh4}, which has
been interpreted as a signature of chemical evolution.
\item[(4)\hspace{2.0mm}]
Both empirical O and Na differential abundance distributions, related to the
whole sample, have been fitted by a straight line with regard to O$_-$, O$_+$,
Na$_-$, Na$_+$ class, respectively, where the slopes related to O$_+$ and
Na$_-$ class were equal and opposite in sign within the errors, while the
contrary was found for O$_-$ and Na$_+$ class, as shown in Table
\ref{t:abs} and in Fig.\,\ref{f:jp12onah4}.   The above results have been
interpreted as consistent with oxygen depletion mainly
turned into sodium enrichment for [O/H]$\ge-1.35$, [Na/H]$\le-1.45$, and
implying oxygen depletion through most preferred channels with respect to Na
for [O/H]$<-1.35$, [Na/H]$>-1.45$.
\item[(5)\hspace{2.0mm}]
In the light of simple MCBR models of chemical evolution in the linear limit,
the ratio of M13 primordial to current mass has been found to be proportional
to the true sodium yield in units of sodium solar abundance, as shown in Table
\ref{t:solin}, where $M_{\rm primordial}/M_{\rm current}=[(\mu_f)_{\rm AF}+
(s_f)_{\rm AF}]/(s_f)_{\rm CE}$.
\end{description}
As current abundance determinations, [Q/H], are affected by large $(\mp1.5$
dex) uncertainties in GC stars, the above results are to be conceived as
mainly qualitative, but the trends shown are expected to be real.   As
already mentioned in Section \ref{disc}, the validity of the proposed
classification may be tested in a twofold manner, namely by extending 
Figs.\,\ref{f:jnrofe} and \ref{f:jnrona} to (i) elements other than Na, Fe,
and (ii) GCs other than M13.


\appendix
\section*{Appendix}

\section{JP12 sample star data}\label{a:data}

The whole set of JP12 sample star data used in the text is listed in Tables
\ref{t:c13m13}-\ref{t:c13m16}, where the following denomination or value
appears on each column: (1) star name; (2) other star name; (3) [O/H]; (4)
[Fe/H]; (5) [Na/H]; (6) $b_{\rm Fe}$ via Eq.\,(\ref{eq:QOH}) where
$a_{\rm Fe}=1.00$; (7) $b_{\rm Na}$ via Eq.\,(\ref{eq:QOH}) where
$a_{\rm Na}=1.25$; (8) population (P, I, E) as defined in the parent paper
(JP12); (9) Fe class according to Eq.\,(\ref{eq:bfe}); (10) Na class according
to Eq.\,(\ref{eq:bna}); (11) Fe-Na class (N, T, H) according to
Eq.\,(\ref{seq:NTH}).
\begin{table}
\caption{M13 star classification according to the parent paper (JP12) and the 
current attempt.   Number abundances, [O/H], [Fe/H], [Na/H], are taken or
inferred from the parent paper.   Intercepts, $b_{\rm Q}$, relate to the 
straight line,  [Q/H] = $a_{\rm Q}$[O/H] + $b_{\rm Q}$, passing through
the point, ([Q/H],[O/H]), where Q = Fe, Na, $a_{\rm Fe}=1.00$,
$a_{\rm Na}=1.25$.   Population P (primitive), I (intermediate), E (extreme),
are defined as in the parent paper.   Classes with different degree of
anomaly, $A_i$, $i=0,\mp1,\mp2,...$, with regard to Q = Fe, Na, are defined as
in the text.   Class $A_{-1}$ is listed as $-A_1$ to save aesthetics.   Class
N (normal), T (transition), H (horizontal branch), are defined as in the text.
See text for further details.}
\label{t:c13m13}
\begin{center}
\begin{tabular}{lllllrrcrrc} \hline
\multicolumn{1}{l}{name} &
\multicolumn{1}{l}{other} &
\multicolumn{1}{c}{[O/H]} &
\multicolumn{1}{c}{[Fe/H]} &
\multicolumn{1}{c}{[Na/H]} &
\multicolumn{1}{c}{$b_{\rm Fe}$} &
\multicolumn{1}{c}{$b_{\rm Na}$} &
\multicolumn{1}{c}{p} &
\multicolumn{1}{c}{Fe} &
\multicolumn{1}{c}{Na} &
\multicolumn{1}{c}{c} \\
\hline
L  324 & V11    & $-$1.96 & $-$1.50 & $-$1.23 & $ $0.46 & $ $1.22 & E & $A_2    $ & $A_3    $ & H \\
L  598 & ...    & $-$1.35 & $-$1.44 & $-$1.58 & $-$0.09 & $ $0.11 & P & $A_1    $ & $A_1    $ & H \\
L  629 & ...    & $-$1.63 & $-$1.57 & $-$1.37 & $ $0.06 & $ $0.67 & I & $A_1    $ & $A_2    $ & H \\
L  194 & II-90  & $-$1.90 & $-$1.49 & $-$1.13 & $ $0.41 & $ $1.25 & E & $A_2    $ & $A_3    $ & H \\
L  973 & I-48   & $-$2.55 & $-$1.50 & $-$1.23 & $ $1.05 & $ $1.96 & E & $A_3    $ & $A_4    $ & H \\
L  835 & V15    & $-$2.16 & $-$1.50 & $-$1.10 & $ $0.66 & $ $1.60 & E & $A_2    $ & $A_3    $ & H \\
L  954 & IV-25  & $-$2.00 & $-$1.48 & $-$1.12 & $ $0.52 & $ $1.38 & E & $A_2    $ & $A_3    $ & H \\
L  940 & ...    & $-$2.09 & $-$1.53 & $-$1.36 & $ $0.56 & $ $1.25 & E & $A_2    $ & $A_3    $ & H \\
L   70 & II-67  & $-$2.45 & $-$1.45 & $-$1.12 & $ $1.00 & $ $1.94 & E & $A_3    $ & $A_4    $ & H \\
L  199 & III-63 & $-$1.51 & $-$1.61 & $-$1.41 & $-$0.10 & $ $0.48 & I & $A_1    $ & $A_1    $ & H \\
L  853 & ...    & $-$1.50 & $-$1.51 & $-$1.31 & $-$0.01 & $ $0.57 & I & $A_1    $ & $A_2    $ & H \\
L  261 & ...    & $-$1.59 & $-$1.59 & $-$1.53 & $ $0.00 & $ $0.46 & I & $A_1    $ & $A_1    $ & H \\
L  262 & ...    & $-$1.52 & $-$1.58 & $-$1.36 & $-$0.06 & $ $0.54 & I & $A_1    $ & $A_2    $ & H \\
L   72 & III-73 & $-$1.45 & $-$1.59 & $-$1.75 & $-$0.14 & $ $0.06 & P & $A_1    $ & $A_1    $ & H \\
L  240 & II-34  & $-$1.82 & $-$1.42 & $-$1.12 & $ $0.40 & $ $1.16 & E & $A_2    $ & $A_3    $ & H \\
L  481 & ...    & $-$2.26 & $-$1.49 & $-$1.12 & $ $0.77 & $ $1.71 & E & $A_2    $ & $A_4    $ & H \\
L  250 & ...    & $-$1.86 & $-$1.58 & $-$1.20 & $ $0.28 & $ $1.12 & E & $A_1    $ & $A_3    $ & H \\
L  384 & ...    & $-$1.25 & $-$1.61 & $-$1.57 & $-$0.36 & $-$0.01 & I & $A_0    $ & $A_1    $ & T \\
L  465 & ...    & $-$1.26 & $-$1.64 & $-$1.79 & $-$0.38 & $-$0.21 & P & $A_0    $ & $A_0    $ & N \\
L   96 & II-76  & $-$1.20 & $-$1.63 & $-$1.72 & $-$0.43 & $-$0.22 & P & $A_0    $ & $A_0    $ & N \\
L  845 & ...    & $-$2.05 & $-$1.57 & $-$1.17 & $ $0.48 & $ $1.39 & E & $A_2    $ & $A_3    $ & H \\
L  745 & I-13   & $-$1.35 & $-$1.55 & $-$1.91 & $-$0.20 & $-$0.22 & P & $A_1    $ & $A_0    $ & T \\
L  584 & ...    & $-$1.40 & $-$1.63 & $-$1.95 & $-$0.23 & $-$0.20 & I & $A_0    $ & $A_0    $ & N \\
L  296 & ...    & $-$1.45 & $-$1.71 & $-$1.43 & $-$0.26 & $ $0.38 & I & $A_0    $ & $A_1    $ & T \\
L  367 & ...    & $-$2.01 & $-$1.57 & $-$1.17 & $ $0.44 & $ $1.34 & E & $A_2    $ & $A_3    $ & H \\
L  316 & III-59 & $-$1.80 & $-$1.62 & $-$1.38 & $ $0.18 & $ $0.87 & E & $A_1    $ & $A_2    $ & H \\
L  549 & ...    & $-$1.35 & $-$1.62 & $-$1.18 & $-$0.27 & $ $0.51 & I & $A_0    $ & $A_2    $ & T \\
L  674 & ...    & $-$1.61 & $-$1.62 & $-$1.18 & $-$0.01 & $ $0.83 & I & $A_1    $ & $A_2    $ & H \\
\hline                                                                      
\end{tabular}                                                               
\end{center}                                                                
\end{table}                                                                 
\begin{table}
\caption{Continuation of Table \ref{t:c13m13}.}
\label{t:c13m14}
\begin{center}
\begin{tabular}{lllllrrcrrc} \hline
\multicolumn{1}{l}{name} &
\multicolumn{1}{l}{other} &
\multicolumn{1}{c}{[O/H]} &
\multicolumn{1}{c}{[Fe/H]} &
\multicolumn{1}{c}{[Na/H]} &
\multicolumn{1}{c}{$b_{\rm Fe}$} &
\multicolumn{1}{c}{$b_{\rm Na}$} &
\multicolumn{1}{c}{p} &
\multicolumn{1}{c}{Fe} &
\multicolumn{1}{c}{Na} &
\multicolumn{1}{c}{c} \\
\hline
L  398 & ...    & $-$1.30 & $-$1.55 & $-$1.47 & $-$0.25 & $ $0.16 & I & $A_0    $ & $A_1    $ & T \\
L  244 & III-52 & $-$1.70 & $-$1.62 & $-$1.36 & $ $0.08 & $ $0.77 & I & $A_1    $ & $A_2    $ & H \\
L  252 & II-33  & $-$1.38 & $-$1.60 & $-$1.58 & $-$0.22 & $ $0.15 & I & $A_0    $ & $A_1    $ & T \\
L  830 & ...    & $-$1.91 & $-$1.59 & $-$1.07 & $ $0.32 & $ $1.32 & E & $A_2    $ & $A_3    $ & H \\
L  938 & IV-53  & $-$1.17 & $-$1.63 & $-$1.21 & $-$0.46 & $ $0.25 & P & $A_0    $ & $A_1    $ & T \\
L  158 & II-57  & $-$1.80 & $-$1.58 & $-$1.52 & $ $0.22 & $ $0.73 & E & $A_1    $ & $A_2    $ & H \\
L  666 & ...    & $-$1.30 & $-$1.57 & $-$1.45 & $-$0.27 & $ $0.17 & I & $A_0    $ & $A_1    $ & T \\
L   77 & III-18 & $-$1.86 & $-$1.62 & $-$1.18 & $ $0.24 & $ $1.14 & E & $A_1    $ & $A_3    $ & H \\
L  169 & III-37 & $-$1.86 & $-$1.52 & $-$1.03 & $ $0.34 & $ $1.29 & E & $A_2    $ & $A_3    $ & H \\
L  825 & ...    & $-$1.61 & $-$1.55 & $-$1.12 & $ $0.06 & $ $0.89 & I & $A_1    $ & $A_2    $ & H \\
L  353 & II-40  & $-$1.23 & $-$1.58 & $-$1.81 & $-$0.35 & $-$0.27 & P & $A_0    $ & $A_0    $ & N \\
L  594 & ...    & $-$1.45 & $-$1.57 & $-$1.12 & $-$0.12 & $ $0.69 & I & $A_1    $ & $A_2    $ & H \\
L  777 & I-24   & $-$1.37 & $-$1.57 & $-$1.43 & $-$0.20 & $ $0.28 & I & $A_1    $ & $A_1    $ & H \\
L 1073 & ...    & $-$1.32 & $-$1.64 & $-$1.58 & $-$0.32 & $ $0.07 & I & $A_0    $ & $A_1    $ & T \\
L  754 & ...    & $-$1.26 & $-$1.60 & $-$1.51 & $-$0.34 & $ $0.07 & I & $A_0    $ & $A_1    $ & T \\
L  198 & ...    & $-$1.51 & $-$1.53 & $-$1.06 & $-$0.02 & $ $0.83 & I & $A_1    $ & $A_2    $ & H \\
L  687 & IV-15  & $-$1.55 & $-$1.65 & $-$1.52 & $-$0.10 & $ $0.42 & I & $A_1    $ & $A_1    $ & H \\
L  863 & I-42   & $-$1.70 & $-$1.59 & $-$1.48 & $ $0.11 & $ $0.64 & I & $A_1    $ & $A_2    $ & H \\
L 1023 & IV-61  & $-$1.71 & $-$1.50 & $-$1.19 & $ $0.21 & $ $0.95 & E & $A_1    $ & $A_2    $ & H \\
L  877 & I-50   & $-$1.45 & $-$1.66 & $-$1.29 & $-$0.21 & $ $0.52 & I & $A_0    $ & $A_2    $ & T \\
L  919 & IV-28  & $-$1.75 & $-$1.62 & $-$1.27 & $ $0.13 & $ $0.92 & I & $A_1    $ & $A_2    $ & H \\
K  656 & ...    & $-$1.31 & $-$1.57 & $-$1.54 & $-$0.26 & $ $0.10 & I & $A_0    $ & $A_1    $ & T \\
L  343 & ...    & $-$1.25 & $-$1.60 & $-$1.41 & $-$0.35 & $ $0.15 & I & $A_0    $ & $A_1    $ & T \\
L  592 & ...    & $-$1.40 & $-$1.68 & $-$1.37 & $-$0.28 & $ $0.38 & I & $A_0    $ & $A_1    $ & T \\
L  476 & ...    & $-$1.32 & $-$1.62 & $-$1.40 & $-$0.30 & $ $0.25 & I & $A_0    $ & $A_1    $ & T \\
L  269 & ...    & $-$1.76 & $-$1.65 & $-$1.44 & $ $0.11 & $ $0.76 & I & $A_1    $ & $A_2    $ & H \\
L  948 & IV-35  & $-$1.39 & $-$1.64 & $-$1.46 & $-$0.25 & $ $0.28 & I & $A_0    $ & $A_1    $ & T \\
L  967 & I-86   & $-$2.01 & $-$1.55 & $-$1.17 & $ $0.46 & $ $1.34 & E & $A_2    $ & $A_3    $ & H \\
L 1030 & I-77   & $-$1.30 & $-$1.61 & $-$1.46 & $-$0.31 & $ $0.17 & I & $A_0    $ & $A_1    $ & T \\
L  644 & ...    & $-$1.66 & $-$1.62 & $-$1.30 & $ $0.04 & $ $0.78 & I & $A_1    $ & $A_2    $ & H \\
L  773 & I-23   & $-$1.35 & $-$1.66 & $-$1.32 & $-$0.31 & $ $0.37 & I & $A_0    $ & $A_1    $ & T \\
L  956 & ...    & $-$1.45 & $-$1.63 & $-$1.26 & $-$0.18 & $ $0.55 & I & $A_1    $ & $A_2    $ & H \\
K  228 & J 3    & $-$1.51 & $-$1.45 & $-$0.93 & $ $0.06 & $ $0.96 & I & $A_1    $ & $A_2    $ & H \\
L  176 & II-87  & $-$1.40 & $-$1.61 & $-$1.24 & $-$0.21 & $ $0.51 & I & $A_0    $ & $A_2    $ & T \\
K  188 & A1     & $-$1.15 & $-$1.52 & $-$1.30 & $-$0.37 & $ $0.14 & I & $A_0    $ & $A_1    $ & T \\
L  436 & ...    & $-$1.70 & $-$1.78 & $-$1.42 & $-$0.08 & $ $0.70 & I & $A_1    $ & $A_2    $ & H \\
L  114 & III-7  & $-$1.45 & $-$1.61 & $-$1.26 & $-$0.16 & $ $0.55 & I & $A_1    $ & $A_2    $ & H \\
\hline                                                                      
\end{tabular}                                                               
\end{center}                                                                
\end{table}                                                                 
\begin{table}
\caption{Continuation of Table \ref{t:c13m14}.}
\label{t:c13m15}
\begin{center}
\begin{tabular}{lllllrrcrrc} \hline
\multicolumn{1}{l}{name} &
\multicolumn{1}{l}{other} &
\multicolumn{1}{c}{[O/H]} &
\multicolumn{1}{c}{[Fe/H]} &
\multicolumn{1}{c}{[Na/H]} &
\multicolumn{1}{c}{$b_{\rm Fe}$} &
\multicolumn{1}{c}{$b_{\rm Na}$} &
\multicolumn{1}{c}{p} &
\multicolumn{1}{c}{Fe} &
\multicolumn{1}{c}{Na} &
\multicolumn{1}{c}{c} \\
\hline
L  370 & ...    & $-$1.85 & $-$1.66 & $-$1.39 & $ $0.19 & $ $0.92 & E & $A_1    $ & $A_2    $ & H \\
L 1043 & BAUM13 & $-$1.43 & $-$1.67 & $-$1.10 & $-$0.24 & $ $0.69 & I & $A_0    $ & $A_2    $ & T \\
L  766 & I-12   & $-$1.65 & $-$1.63 & $-$1.43 & $ $0.02 & $ $0.63 & I & $A_1    $ & $A_2    $ & H \\
L  172 & III-45 & $-$1.40 & $-$1.58 & $-$1.41 & $-$0.18 & $ $0.34 & I & $A_1    $ & $A_1    $ & H \\
L   26 & J38    & $-$1.65 & $-$1.51 & $-$1.23 & $ $0.14 & $ $0.83 & I & $A_1    $ & $A_2    $ & H \\
L  168 & II-28  & $-$1.03 & $-$1.64 & $-$2.30 & $-$0.61 & $-$1.01 & P & $A_0    $ &$-A_1    $ & N \\
L  193 & II-94  & $-$1.60 & $-$1.56 & $-$1.01 & $ $0.04 & $ $0.99 & I & $A_1    $ & $A_2    $ & H \\
L  726 & IV-19  & $-$1.28 & $-$1.54 & $-$1.38 & $-$0.26 & $ $0.22 & I & $A_0    $ & $A_1    $ & T \\
L  793 & ...    & $-$1.05 & $-$1.59 & $-$1.72 & $-$0.54 & $-$0.41 & P & $A_0    $ & $A_0    $ & N \\
K  699 & X 24   & $-$1.38 & $-$1.60 & $-$1.20 & $-$0.22 & $ $0.53 & I & $A_0    $ & $A_2    $ & T \\
L  677 & IV-4   & $-$1.65 & $-$1.59 & $-$0.88 & $ $0.06 & $ $1.18 & I & $A_1    $ & $A_3    $ & H \\
L   18 & ...    & $-$1.55 & $-$1.65 & $-$1.54 & $-$0.10 & $ $0.40 & I & $A_1    $ & $A_1    $ & H \\
L  800 & IV-18  & $-$1.30 & $-$1.59 & $-$1.35 & $-$0.29 & $ $0.27 & I & $A_0    $ & $A_1    $ & T \\
L 1032 & I-76   & $-$1.40 & $-$1.59 & $-$1.67 & $-$0.19 & $ $0.08 & E & $A_1    $ & $A_1    $ & H \\
L  871 & I-19   & $-$2.46 & $-$1.62 & $-$0.99 & $ $0.84 & $ $2.09 & E & $A_3    $ & $A_4    $ & H \\
L  955 & IV-22  & $-$1.45 & $-$1.54 & $-$1.25 & $-$0.09 & $ $0.56 & I & $A_1    $ & $A_2    $ & H \\
L  609 & ...    & $-$1.30 & $-$1.69 & $-$1.22 & $-$0.39 & $ $0.41 & I & $A_0    $ & $A_1    $ & T \\
L   81 & II-23  & $-$1.58 & $-$1.60 & $-$1.04 & $-$0.02 & $ $0.93 & I & $A_1    $ & $A_2    $ & H \\
L 1001 & I-49   & $-$1.30 & $-$1.54 & $-$1.60 & $-$0.24 & $ $0.02 & P & $A_0    $ & $A_1    $ & T \\
K  422 & ...    & $-$1.10 & $-$1.54 & $-$1.32 & $-$0.44 & $ $0.05 & I & $A_0    $ & $A_1    $ & T \\
L 1060 & I-65   & $-$1.65 & $-$1.57 & $-$1.36 & $ $0.08 & $ $0.70 & I & $A_1    $ & $A_2    $ & H \\
L  162 & III-43 & $-$1.56 & $-$1.48 & $-$1.40 & $ $0.08 & $ $0.55 & I & $A_1    $ & $A_2    $ & H \\
L  488 & ...    & $-$1.35 & $-$1.55 & $-$1.30 & $-$0.20 & $ $0.39 & I & $A_1    $ & $A_1    $ & H \\
L 1051 & IV-78  & $-$1.86 & $-$1.57 & $-$1.20 & $ $0.29 & $ $1.12 & E & $A_1    $ & $A_3    $ & H \\
L 1114 & ...    & $-$1.35 & $-$1.65 & $-$1.64 & $-$0.30 & $ $0.05 & I & $A_0    $ & $A_1    $ & T \\
L  557 & ...    & $-$1.13 & $-$1.58 & $-$1.79 & $-$0.45 & $-$0.38 & P & $A_0    $ & $A_0    $ & N \\
L   79 & ...    & $-$1.00 & $-$1.55 & $-$1.81 & $-$0.55 & $-$0.56 & P & $A_0    $ & $A_0    $ & N \\
K  659 & ...    & $-$1.40 & $-$1.39 & $-$1.06 & $ $0.01 & $ $0.69 & I & $A_1    $ & $A_2    $ & H \\
L  140 & III-25 & $-$1.50 & $-$1.61 & $-$1.00 & $-$0.11 & $ $0.88 & I & $A_1    $ & $A_2    $ & H \\
K  674 & ...    & $-$0.90 & $-$1.59 & $-$1.33 & $-$0.69 & $-$0.20 & I & $A_0    $ & $A_0    $ & N \\
L  787 & I-2    & $-$1.28 & $-$1.58 & $-$1.11 & $-$0.30 & $ $0.49 & I & $A_0    $ & $A_1    $ & T \\
L 1050 & ...    & $-$1.40 & $-$1.47 & $.$.... & $-$0.07 & $ $.... & . & $A_1    $ & $...    $ & . \\
L 1072 & IV-80  & $-$1.45 & $-$1.55 & $-$0.91 & $-$0.10 & $ $0.90 & I & $A_1    $ & $A_2    $ & H \\
L  423 & II-7   & $-$1.00 & $-$1.51 & $-$1.62 & $-$0.51 & $-$0.37 & P & $A_0    $ & $A_0    $ & N \\
L  824 & I-39   & $-$1.25 & $-$1.56 & $-$1.19 & $-$0.31 & $ $0.37 & I & $A_0    $ & $A_1    $ & T \\
L 1096 & I-67   & $-$1.65 & $-$1.58 & $-$1.25 & $ $0.07 & $ $0.81 & I & $A_1    $ & $A_2    $ & H \\
L 1097 & ...    & $-$1.80 & $-$1.48 & $-$1.28 & $ $0.32 & $ $0.97 & E & $A_2    $ & $A_2    $ & H \\
\hline                                                                      
\end{tabular}                                                               
\end{center}                                                                
\end{table}                                                                 
\begin{table}
\caption{Continuation of Table \ref{t:c13m15}.}
\label{t:c13m16}
\begin{center}
\begin{tabular}{lllllrrcrrc} \hline
\multicolumn{1}{l}{name} &
\multicolumn{1}{l}{other} &
\multicolumn{1}{c}{[O/H]} &
\multicolumn{1}{c}{[Fe/H]} &
\multicolumn{1}{c}{[Na/H]} &
\multicolumn{1}{c}{$b_{\rm Fe}$} &
\multicolumn{1}{c}{$b_{\rm Na}$} &
\multicolumn{1}{c}{p} &
\multicolumn{1}{c}{Fe} &
\multicolumn{1}{c}{Na} &
\multicolumn{1}{c}{c} \\
\hline
L   29 & II-63  & $.$.... & $-$1.59 & $-$1.07 & $.$.... & $ $.... & . & $...    $ & $...    $ & . \\
K  224 & J 37   & $-$1.15 & $-$1.53 & $-$1.03 & $-$0.38 & $ $0.41 & I & $A_0    $ & $A_1    $ & T \\
L  137 & ...    & $-$1.25 & $-$1.47 & $-$1.07 & $-$0.22 & $ $0.49 & I & $A_0    $ & $A_1    $ & T \\
L   16 & J50    & $-$1.17 & $-$1.61 & $-$1.79 & $-$0.44 & $-$0.33 & P & $A_0    $ & $A_0    $ & N \\
K  647 & ...    & $-$1.65 & $-$1.58 & $-$1.51 & $ $0.07 & $ $0.55 & I & $A_1    $ & $A_2    $ & H \\
L   93 & III-40 & $-$1.00 & $-$1.42 & $-$1.03 & $-$0.42 & $ $0.22 & I & $A_0    $ & $A_1    $ & T \\
L 1095 & I-69   & $-$1.30 & $-$1.62 & $-$1.68 & $-$0.32 & $-$0.05 & P & $A_0    $ & $A_1    $ & T \\
L    6 & J11    & $-$1.70 & $-$1.68 & $-$1.32 & $ $0.02 & $ $0.81 & I & $A_1    $ & $A_2    $ & H \\
L  101 & II-60  & $-$1.28 & $-$1.68 & $-$2.00 & $-$0.40 & $-$0.40 & P & $A_0    $ & $A_0    $ & N \\
L   32 & II-64  & $-$1.43 & $-$1.44 & $-$0.94 & $-$0.01 & $ $0.85 & I & $A_1    $ & $A_2    $ & H \\
CM  12 & ...    & $-$0.65 & $-$1.39 & $-$1.40 & $-$0.74 & $-$0.59 & P &$-A_1    $ & $A_0    $ & N \\
\hline                                                                      
\end{tabular}                                                               
\end{center}                                                                
\end{table}                                                                 

\section{Differential element abundance distribution due to cosmic scatter}
\label{a:cosca}

Let a sample be made of $N$ long-lived coeval stars belonging to the same
generation, where the abundance of the primordial gas was affected from cosmic
scatter.   Let the related distribution be Gaussian in [Q/H] with respect to a
selected element, Q, and normalized to the sample population, $N$.
The explicit expression reads:
\begin{lefteqnarray}
\label{eq:gaus}
&& \frac{\diff N}N=\frac1{\sqrt{2\pi}\sigma}\exp\left(-\frac{x^2}{2\sigma^2}
\right)\diff x~~; \\
\label{eq:erQ}
&& x=[{\rm Q/H}]-<[{\rm Q/H}]>~~;
\end{lefteqnarray}
where $\diff N$ is the expected star number within a bin, centered on $x$, of
infinitesimal width, $\diff x$.

Keeping in mind [Q/H]$=\log\phi_{\rm Q}$ to a good extent (e.g., Caimmi
2011a), the infinitesimal bin width reads:
\begin{lefteqnarray}
\label{eq:dx}
&& \diff x=\diff{\rm [Q/H]}=\frac1{\ln10}\frac{\diff\phi_{\rm Q}}
{\phi_{\rm Q}}~~;
\end{lefteqnarray}
which is equivalent to:
\begin{lefteqnarray}
\label{eq:dxdfi}
&& \frac{\diff x}{\diff\phi_{\rm Q}}=\frac1{\ln10}\frac1{\phi_{\rm Q}}~~;
\end{lefteqnarray}
according to Eq.\,(\ref{eq:erQ}).

The theoretical differential element abundance distribution due to cosmic
scatter is inferred from the theoretical counterpart of Eq.\,(\ref{eq:psie}):
\begin{lefteqnarray}
\label{eq:psit}
&& \psi_{\rm Q}=\log\frac{\diff N}{N\diff\phi_{\rm Q}}~~;
\end{lefteqnarray}
after substitution of Eqs.\,(\ref{eq:gaus}) and (\ref{eq:dxdfi}) into
(\ref{eq:psit}).   The result is:
\begin{lefteqnarray}
\label{eq:psics}
&& (\psi_{\rm Q})_{\rm cs}=\log\left\{\frac1{\ln10}\frac1{\sqrt{2\pi}\sigma}
\exp\left[-\frac{(\log\phi_{\rm Q}-\overline{\log\phi_{\rm Q}})^2}{2\sigma^2}
\right]\frac1{\phi_{\rm Q}}\right\}~~;
\end{lefteqnarray}
where cs denotes cosmic scatter and $\overline{\log\phi_{\rm Q}}=\,
<[{\rm Q/H}]>$.

\end{document}